\documentclass{aa}  %

\usepackage{graphicx}
\usepackage{txfonts,textcomp}
\usepackage[group-separator={,}]{siunitx}
\usepackage{hyperref}
\begin{document}

   \title{Planetary nebulae of the Large Magellanic Cloud 
\\ I. A multiwavelength analysis}

%  \subtitle{I. Gas }

   \author{S. Tosi\inst{1,2,3}, F. Dell'Agli\inst{2}, D. Kamath\inst{2,4}, L. Stanghellini\inst{5}, 
   P. Ventura\inst{2,6}, S. Bianchi\inst{1}, M. A. Gómez-Muñoz\inst{7,8}, D. A. Garc{\'\i}a-Hern{\'a}ndez\inst{7,8}}

   \institute{Dipartimento di Matematica e Fisica, Università degli Studi Roma Tre, 
              via della Vasca Navale 84, 00100, Roma, Italy \and
              INAF, Observatory of Rome, Via Frascati 33, 00077 Monte Porzio Catone (RM), Italy \and 
              LNF, Laboratori Nazionali Fascati, Via Enrico Fermi, 54, 00044 Frascati Roma, Italy \and
              School of Mathematical and Physical Sciences, Macquarie University, Balaclava Road, Sydney, NSW 2109, Australia \and
              NSF's NOIRLab, 950 Cherry Ave., Tucson, AZ 85719, USA \and
              Istituto Nazionale di Fisica Nucleare, section of Perugia, Via A. Pascoli snc, 06123 Perugia, Italy \and
              Instituto de Astrof\'isica de Canarias (IAC), E-38205 La Laguna, Tenerife, Spain \and
              Departamento de Astrof\'isica, Universidad de La Laguna (ULL), E-38206 La Laguna, Tenerife, Spain
              }

   \date{Received September 15, 1996; accepted March 16, 1997}

% \abstract{}{}{}{}{} 
% 5 {} token are mandatory

 \abstract
 {Planetary nebulae (PNe) have three main components: a central star (CS), ionized gas, and dust in the nebula. Each contains critical chemical fingerprints of the PN's evolution, which serve as tracers of the evolution, nucleosynthesis, and dust production that occurred during the preceding asymptotic giant branch (AGB) phase.  }
   {We aim to build a bridge to link the PN phase to the evolution of progenitors, to better understand the dust production and mass-loss mechanism during the final AGB phase. Here we present a comprehensive study of nine Large Magellanic Cloud spherical or elliptical PNe whose observations from the UV through the IR are available in the literature. We characterize nebulae and CSs, finding information necessary to reconstruct the evolutionary history of mass-loss and dust production, such as as the amount of gas that makes up the nebula and the dust that surrounds the CS. }
   {We compared the observed energy distribution of the selected PNe to that obtained from photoionization modeling, taking the presence of dust  into account. The physical and chemical parameters of the CSs were then compared with predictions from the evolutionary tracks.}
   {We characterize the source, assigning a progenitor, early-AGB mass to each CS. We estimate the mass of the nebula and the dust-to-gas ratio. For five objects, we find evidence for the presence of a near-IR bump, which would indicate the presence of hot dust. }
   {}


   \titlerunning{Characterizing planetary nebulae of the Large Magellanic Cloud}
   \authorrunning{Tosi et al.}
   \maketitle
%
%-------------------------------------------------------------------

\section{Introduction}
The final evolutionary phase of stars with masses between $\sim$\,1 and 8$\,\rm{M}_\odot$, that is to say, low- and intermediate-mass stars (LIMSs), is a crucial aspect in understanding the environmental history of their formation sites. As these stars exhaust the helium in their core they ascend the asymptotic giant branch (AGB), expanding their envelope and reaching high luminosities of up to about 10$^{5}\,\rm L_{\odot}$ at cool temperatures of ~3\,000\,K or lower.

During the AGB phase, several processes alter the surface chemical composition of the stars \citep{karakas14,ventura22}. The third dredge-up \citep[TDU;][]{iben74}, a process involving the deepening of the convective envelope in regions previously influenced by internal mixing and nucleosynthesis, increases the surface abundance of $^{12}$C in stars with masses lower than $4\rm{M}_\odot$ (depending on the initial metallicity). Repeated TDU episodes can eventually result in the formation of carbon stars, characterized by a surface carbon-to-oxygen ratio  (C/O) greater than unity. In the intermediate-mass range, the activation of proton capture nucleosynthesis at the base of the convective envelope, referred to as hot bottom burning \citep[HBB;][]{sackmann92}, primarily depletes $^{12}$C and synthesizes $^{14}$N, thus inhibiting the formation of carbon stars. Toward the end of the AGB phase, stars shed their expanded and cool envelopes, releasing gas into the surrounding medium. This gas reflects the effects of processes such as TDU and HBB on the star's surface chemistry.  The expelled material is carried away by a cool and dense stellar wind, providing an optimal environment for the production of dust \citep{habing96,gail2009}. As a result, AGB stars are significant sources of dust in their host galaxies, including the Magellanic Clouds \citep[e.g.,][]{boyer12,matsuura11,matsuura13,raffa14,raffa23}.
Stars that evolve as oxygen-rich\footnote{ We note that in this paper, “carbon-rich” and “oxygen-rich” refer to the nebular gas being C/O>1 and C/O<1, while CRD and ORD refer to carbon- or oxygen-rich nature of the dust.} primarily produce silicate dust, while those transitioning to the carbon-rich stage mainly produce carbonaceous dust. Evidence suggests a gradual increase in dust production during the AGB phase, with a peak toward its conclusion \citep{flavia15,marini21,ester23}. When the majority of their circumstellar envelope is lost, LIMSs begin their contraction toward the planetary nebula (PN) stage. During the PN phase, the central star (CS) temperature increases to more than 100\,000\,K, while the residual dust expands and cools. Importantly, there are no further significant processes altering the surface chemical composition of the stars as they transition toward the PN phase.

Several studies have demonstrated the utility of comparing data from photometry and spectroscopy with predictions from evolutionary models to characterize the main physical and chemical properties of evolved stars. This approach aids in reconstructing important aspects of the earlier evolutionary stages of the stars.
Notably, studies of post-AGB stars have played a crucial role in providing new insights into mass-loss rates and dust production as a function of initial mass and metallicity \citep{tosi22, tosi23, flavia23}. 
The PN phase also has significant potential, providing a complementary perspective for studying LIMSs in their evolved stages. By comparing PN chemical compositions with predictions from stellar evolution models, it is possible to reconstruct the AGB history, including nucleosynthesis and mixing processes \citep[][]{devika23, ventura15, ventura16, ventura17, garciarojas18, stanghellini22}. Moreover, \citet{flavia23b} demonstrate the feasibility of reconstructing final mass-loss episodes and dust production during the AGB phase through the analysis of spectral energy distributions (SEDs) of PNe. 

This study aims to expand upon the methodology introduced by \citet{flavia23b}, who  applied it to a single PN, by applying it to a broader sample of PNe in the Large Magellanic Cloud (LMC). Several studies have explored various spectral regions to uncover fundamental PN properties in the LMC. These investigations have addressed the morphology and ionization of the nebula \citep{shaw01, shaw06, stanghellini99, stanghellini02, stanghellini05}, the abundance distributions \citep{leisy06}, the characteristic of the CSs \citep[hereinafter EV03]{villaver03}, and the presence of dust \citep[hereinafter LS07]{stanghellini07}. Furthermore, \citet[][hereinafter PV15]{ventura15} investigated a subset of LMC PNe that likely evolved from single stars, offering insights into the progenitor mass during the AGB phase by studying the observed carbon and oxygen abundances.  

While studies have so far focused on specific PN properties, our study aims to splice the observed spectra and photometry, from the IR to the UV, and use this to univocally interpret each nebula plus CS system, including the dust. This goal consists of assigning with reasonable certainty an initial mass and metallicity to each observed system, under the assumption of single-star evolution. Leveraging the well-constrained distance of the LMC, we utilized it as an ideal laboratory for precise determinations of luminosities and initial stellar masses. We did so using tailored photoionization modeling. This approach enabled us to characterize the physical properties of the CS, nebular gas, and dust of each PN in the sample. Additionally, we compared the physical characteristics (e.g., the luminosity and effective temperature of the star, as well as the gas mass of the nebula) and chemical characteristics of PNe with a wide set of ATON stellar evolution models \citep{ventura98} specifically extended to the PN phase. This comparative analysis enabled a robust determination of the progenitor mass of the CS, thereby establishing the connection between observed PNe and their past evolutionary history.

The paper is structured as follows: in Sect. \ref{samplemeth} we describe the selected PN sample and the methodology used to reproduce the SED of each source. The results of the SED analysis, which involved comparing observations with synthetic modeling, are presented in Sect. \ref{SED}. In Sect. \ref{aton} we present a series of stellar evolutionary tracks computed to interpret the observed CS luminosity, effective temperature, and nebular abundances. The evolutionary history of each PN, including the determination of the progenitor mass, is presented in Sect. \ref{mass}. Finally, in Sect. \ref{dust} we discuss the dusty features, utilizing the metallicity and abundances determined in the previous sections. 

\section{The sample and the methodology}
\label{samplemeth}

\subsection{Observational data}
\label{obs}

Our study relied on a selected sample of LMC PNe, whose absolute nebular and CS parameters benefit from the accurate distances to the sources. We specifically selected sources with observed mid-IR spectra to directly assess the presence of dust within the nebula. Additionally, UV spectra and photometric data spanning from UV to optical and IR wavelengths were collected to construct complete SEDs. The final sample of nine LMC PNe comprises round and elliptical PNe, a selection made to exclude close binary systems that might exhibit significant deviations from single-star evolution. This choice aligns with the spherical symmetry assumption of the photoionization model used in our analysis.

\begin{figure*}
\vskip-40pt
\begin{minipage}{0.46\textwidth}
\resizebox{1.\hsize}{!}{\includegraphics{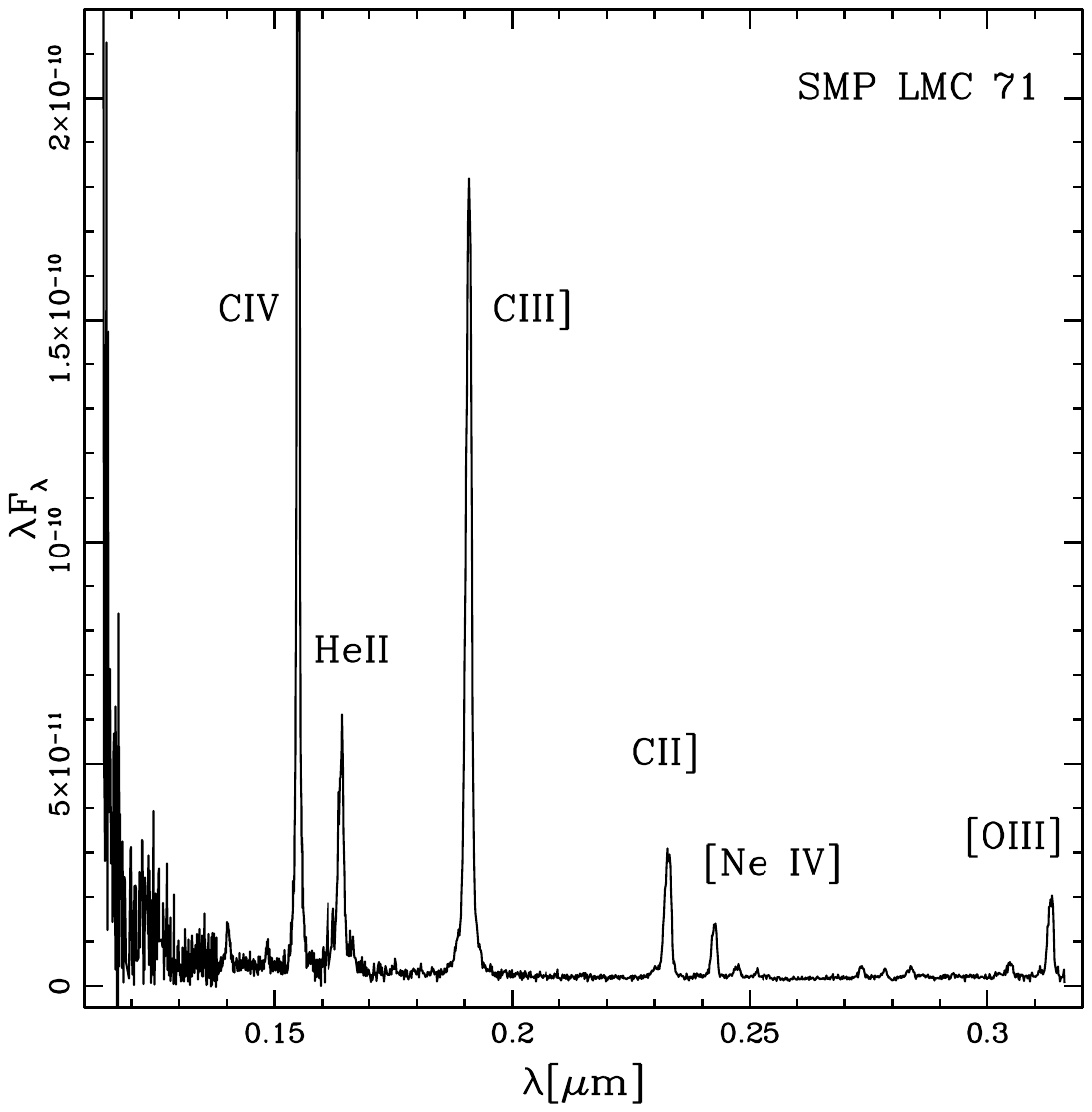}}
\end{minipage}
\begin{minipage}{0.46\textwidth}
\resizebox{1.\hsize}{!}{\includegraphics{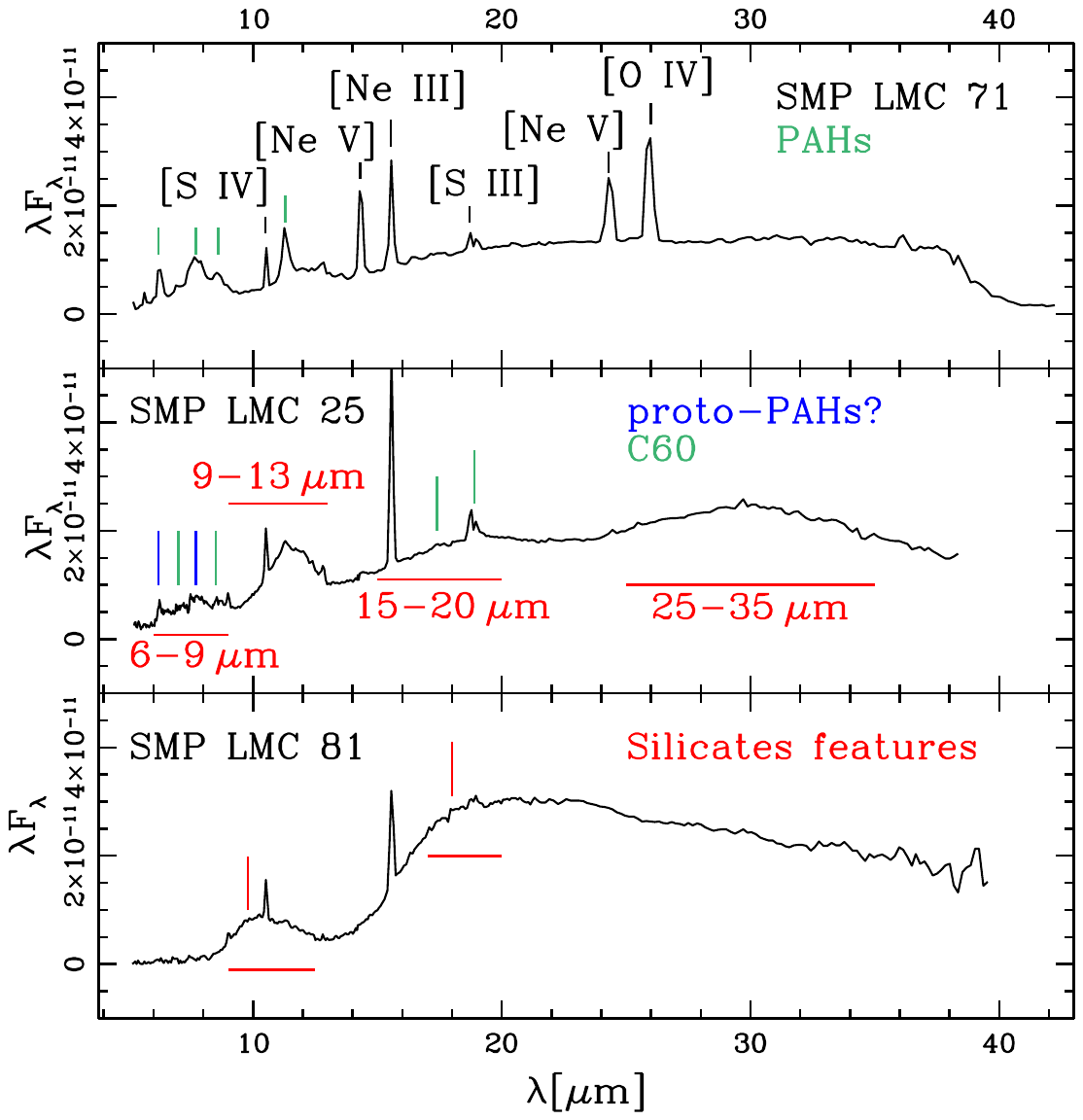}}
\end{minipage}
\vskip-60pt
\caption{Atomic lines and dust features that mainly characterize the spectra of the PNe. Left: Atomic lines in the HST/STIS spectrum of SMP LMC 71. Right: Mid-IR \textit{Spitzer}/IRS spectra of SMP LMC 71 (top), SMP LMC 25 (middle), and SMP LMC 81 (bottom). The upper right panel shows the atomic nebular emission lines and the classical PAH features, while the middle and lower panels display the dust features typically observed in LMC CRD and ORD PNe, respectively.}
\label{ffeatures}
\end{figure*}

The names and coordinates of our targets are given in Table \ref{tabsample}, alongside their morphology, derived from \textit{Hubble} Space Telescope (HST) images \citep{shaw01, shaw06, stanghellini99, stanghellini02}. 
The photometric data were sourced from various catalogs: the UBV and the IRAC data from \cite{reid14} and \cite{lasker08} and WISE photometry from  \cite{cutri12}. The mid-IR spectra are obtained from the \textit{Spitzer} Infrared Spectrograph (IRS; LS07) and UV spectra from the HST/Space Telescope Imaging Spectrograph \citep[STIS;][hereinafter LS05]{stanghellini05}. The typical lines and features visible in these spectra are highlighted in Fig. \ref{ffeatures}. All UV spectra and photometry are corrected for Galactic foreground and LMC extinction using the procedure described in LS05. Furthermore, where available, abundance measurements of C and O from LS05, \cite{leisy06} and \citet{henry89} have been included in Table \ref{tabinput}.

\begin{table*}[ht]
\caption{ID, coordinates, and morphology of the PNe studied in the present work.}
\label{tabsample}      
\centering
%\addtolength{\leftskip}{+2cm}
\addtolength{\leftskip}{-1cm}
\begin{tabular}{c c c c }    
\hline 
ID   &  R.A. (J2000.0)   &  Decl. (J2000.0)  & Morphology$^{a,b,c,d}$     \\ 
\hline
\\
SMP LMC 4   &  04 43 21.50 &  -71 30 09.5  & Elliptical \\
SMP LMC 18  &  05 03 42.64 &  -70 06 47.8  & Round  \\
SMP LMC 25  &  05 06 24.00 &  -69 03 19.2  & Round \\
SMP LMC 34  &  05 10 17.18 &  -68 48 23.0  & Elliptical \\
SMP LMC 66  &  05 28 41.20 &  -67 33 39.0  & Elliptical  \\
SMP LMC 71  &  05 30 33.22 &  -70 44 38.4  & Elliptical \\
SMP LMC 80  &  05 34 38.87 &  -70 19 56.9  & Round \\
SMP LMC 81  &  05 35 20.92 &  -73 55 30.1  & Round \\
SMP LMC 102 &  06 29 32.93 &  -68 03 32.9  & Round \\
\hline 
\label{tabsample}
\end{tabular} 
\\
\textbf{Notes:} $^a$ Shaw et al.  2001;
$^b$ Shaw et al. 2006;
$^c$ Stanghellini et al. 1999; 
$^d$ Stanghellini et al. 2002;  \\
\end{table*}

\begin{table*}
\caption{Input parameters used for the spectral synthesis code CLOUDY. }      
\centering
%\addtolength{\leftskip}{+2cm}
\addtolength{\leftskip}{-1cm}
\begin{tabular}{c c c c c c c }    
\hline 
ID   &  log(C/H)+12   &  log(O/H)+12  &  $\rm{n}_{\rm{H}}$ [m$^{-3}$] & log($\rm{R}_{\rm{in}}/[\rm{m}]$)     \\ 
\hline
\\ 
SMP LMC 4    &  8.66$^a$ $^{+0.1}_{-0.1}$   &  8.61$^b$   $^{+0.1}_{-0.1}$       &  $3.55\times 10^{-3}$  &  15.56   \\ \\ 
SMP LMC 18   &  8.37$^a$ $^{+0.1}_{-0.1}$   &  $\sim$ 7.87$^b$ $^{+0.1}_{-0.1}$  &  $3.31\times 10^{-3}$  &  15.42   \\ \\
SMP LMC 25   &  8.29$^a$ $^{+0.1}_{-0.1}$   &  8.17$^b$  $^{+0.1}_{-0.1}$        &  $1.51\times 10^{-2}$  &  15.08   \\ \\
SMP LMC 34   &  8.13$^a$ $^{+0.1}_{-0.1}$   &  8.46$^c$  $^{+0.4}_{-0.4}$        &  $4.79\times 10^{-3}$  &  15.22   \\ \\
SMP LMC 66   &  8.51 $^b$   $^{+0.1}_{-0.1}$  &  8.31$^b$  $^{+0.1}_{-0.1}$      &  $2.51\times 10^{-3}$  &  15.45   \\ \\
SMP LMC 71   &  8.90$^a$  $^{+0.1}_{-0.1}$  &  8.63$^b$  $^{+0.1}_{-0.1}$        &  $1.10\times 10^{-2}$  &  15.21   \\ \\
SMP LMC 80   &  7.51$^a$  $^{+0.1}_{-0.1}$  &  8.34$^b$  $^{+0.1}_{-0.1}$        &  $7.94\times 10^{-3}$  &  15.04   \\ \\
SMP LMC 81   &  7.16$^a$  $^{+0.1}_{-0.1}$  &  8.25$^b$  $^{+0.1}_{-0.1}$        &  $1.26\times 10^{-3}$  &  14.99   \\ \\
SMP LMC 102  &  8.65$^a$  $^{+0.1}_{-0.1}$  &  8.28$^b$  $^{+0.1}_{-0.1}$        &  $9.00\times 10^{-4}$  &  15.57   \\

\hline 
\label{tabinput}
\end{tabular}\\
\textbf{Notes.} In the colums are reported: source ID, observed abundances of carbon and oxygen, electron density, and the inner radius of the nebulae.
$^a$  Stanghellini et al. 2005;
$^b$ Leisy \& Dennefeld 2006;
$^c$ Henry, Liebert \& Boroson 1989 \\
\end{table*}

\subsection{The photoionization modeling}
\label{input}

The aim of this analysis is to reproduce the UV to mid-IR photometric and spectroscopic data of nine LMC PNe. Specifically, for each target PN, we aim to determine the luminosity and effective temperature of the CS, the gas mass of the nebula, the chemical composition and temperature of the dust and the dust-to-gas ratio. We generated synthetic spectra using the spectral synthesis code CLOUDY \citep[v22.02; ][]{ferland17}, which includes photoionization and radiative transfer calculations designed to simulate gaseous physical conditions.

The input parameters for the synthetic spectra are reported in Table \ref{tabinput}. Below, we describe in detail the assumptions of our SED modeling concerning the CSs (1), the nebulae (2), and dust emission (3).

(1) To reproduce the CS emission, we used a grid of nonlocal thermodynamic equilibrium (non-LTE), line-blanketed, plane-parallel, hydrostatic models of atmospheres from \cite{rauch03}. For SMP LMC 34 and SMP LMC 18, which are not in the \citet{rauch03} grid due to their effective temperatures being $\lesssim$ 50\,000\,K, we used the non-LTE, line-blanketed, and wind-blanketed model of \cite{pauldrach01}. For the gas chemical composition, we applied the \cite{aller83} and \cite{khromov89} schematization, scaling the abundances to achieve agreement with the C and O from LS05, \cite{leisy06}, and \citet[see our Table \ref{tabinput}]{henry89}.

(2) We modeled the nebulae with a spherical shape, consistent with the morphology of the selected PNe (see Sect. \ref{obs}). 
We used a constant hydrogen density $\rm{n}_{\rm{H}}$ (see Table \ref{tabinput}), chosen to reproduce an electron density $\rm{N}_e$ in agreement with the values reported in LS05. We modeled the PNe to achieve an electron temperature $\rm{T}_e$ consistent with the value listed in LS05 (see Appendix \ref{appendix}). We also imposed the PN  inner radii R$_{\rm{in}}$ reported in Table \ref{tabinput}. We derived R$_{\rm{in}}$ using the ionization schematization given in \cite{osterbrok06}, assuming that it is 0.7 times R$_{\rm{phot}}$, a parameter associated with the observed radii of [OIII] \citep{shaw01}. 
    
(3) We considered different dust species and selected the ones able to better reproduce the IR excess and dust and molecular features visible in the \textit{Spitzer} spectra (see Appendix \ref{appendix}). To model the IR emission of PNe that show evidence of carbon-rich dust (CRD), we considered the presence of amorphous carbon, silicon carbide, graphite, and polycyclic aromatic hydrocarbons (PAHs), while for the PNe with oxygen-rich dust (ORD) we used amorphous silicates. As claimed in \citet{stanghellini22}, the CRD sources usually correspond to the condition C/O>1, while the ORD ones generally correspond to C/O<1. For the present work we also considered possible exceptions, where both silicates and carbonaceous dust are present in the nebula (mixed chemistry). This possibility is particularly relevant when the C/O ratio is uncertain (e.g., for the SMP LMC 34).

For the different dust species, we used the CLOUDY built-in optical constants, namely the real and imaginary parts of the complex refractive index, n $+$ ik, from \citet{roleau91} for the amorphous carbon, \citet{laor93} for silicon carbide, \citet{ferland17} for PAHs, and \citet{martin91} for graphite and amorphous silicates.  The grain size distribution used to model the CRD and the ORD PNe is the interstellar medium one described in \citet{mathis77}, which ranges from 0.005 to 0.25$\,\mu$m. The only exception are PAHs, for which the scheme of \citet{abel08} and the 15C atoms described in CLOUDY, ranging from 0.0004$-$0.0011\,$\mu$m, is applied. 

We iteratively adjusted the parameters until finding the combination that best matches the synthetic spectrum with the observed photometry and spectra (HST and \textit{Spitzer}). In this process, we took the continua, the atomic lines, and the dust features that characterize the spectra into account. The comparison between synthetic and HST/STIS spectra, and photometric data with synthetic photometry in the UV and optical wavelengths allowed us to determine the values of luminosity and effective temperature.  A comparison between the effective temperature derived in the present work and the one obtained using the Zanstra method \citep{zanstra1931} in EV03 is presented in Appendix \ref{appendix}. 

From the SED analysis, particularly from the distribution of the mid-IR photometry, we derived the thickness of the nebula, $\Delta$R, which is directly linked to the nebular gaseous mass, $\rm{M}_{\rm{gas}}$. We note that from the sum R$_{\rm{in}}$+$\Delta$R we obtain a value that is in agreement within the 25\% with the photometric radius of the nebulae \citep{shaw01}. Additionally, we estimated the dust temperature (see Table \ref{tab1}) self-consistently using the photoionization code. This setting produces an IR excess in the synthetic SED that is in agreement with the observed one. The only exception is SMP LMC 4, for which we find that to reproduce the IR excess observed is necessary a higher dust temperature (see Sect. \ref{SED}). We also estimated the dust-to-gas mass ratios $\delta$ for the different dust species, which can be used as proxies for the amount of dust present in the PNe.  

To estimate uncertainties, we systematically varied each parameter while keeping the others fixed until the synthetic photometry deviated by 20\% from the observed values (see Appendix \ref{appendix} for a detailed comparison between observations and models for each PN). The photoionization models are computed to reproduce the entire spectrum from the UV to the IR, constraining the contribution from all three main components of the PN: CS, nebula, and dust. This comprehensive approach is much more challenging compared to studies that typically focus on isolated segments of the spectrum  \citep{barria18, toala21}.

\begin{table*}
\caption{Stellar parameter, dust properties, and nebula characteristics derived in the present analysis.  }
\label{tab1}      
\centering
%\addtolength{\leftskip}{+2cm}
\addtolength{\leftskip}{-0.5cm}
\begin{tabular}{c c c c c c c c c}    
\hline 
ID             &  L$/\rm{L}_\odot$            &  $\rm{T}_{\rm{eff}}$[K]            &  log($\Delta$R/[\rm{m}])          & $\rm{M}_{\rm{gas}}$$/~\rm{M}_\odot$    &  log($\delta$[C])                           &  log($\delta$[PAHs])  & $\rm{T}_d$[C] [K]     &  $\rm{T}_d$[PAHs] [K]                 \\ 
\hline
  CRD 
\\
\hline
\\
SMP LMC 4    &   6\,500 $^{+600 }  _{-100  } $   &   105\,000 $^{+5\,000 }_{-5\,000 }  $  &    13.68 $ ^{+0.03 } _{-0.01 } $ &   0.034 $^{+0.002} _{-0.001}$ &  -2.20  $^{+0.07}_{-0.12}$   &  $-$                        &  85  $^{+20}_{-19} $ & $  -  $\\ \\
SMP LMC 18   &  2\,000  $^{+700 }  _{-500  } $   &   50\,000  $^{+6\,000 }_{-5\,000 }  $  &    14.37 $ ^{+0.08 } _{-0.08 } $ &   0.086 $^{+0.020} _{-0.016}$ &   -3.61 $^{+0.38} _{-0.01}$  &  -5.04 $^{+0.22} _{-0.38}$  &  103 $^{+27}_{-25} $ & $181^{+10}_{-10} $ \\ \\
SMP LMC 25   &  4\,900  $^{+400 }  _{-200  } $   &   60\,000  $^{+10\,000}_{-10\,000}  $  &    14.45 $ ^{+0.01 } _{-0.01 } $ &   0.110 $^{+0.003} _{-0.003}$ &   -2.58 $^{+0.15} _{-1.17}$  &  -3.77 $^{+0.30} _{-0.10}$  &  134 $^{+32}_{-30} $ & $227^{+10}_{-10} $ \\ \\
SMP LMC 34   &  4\,500  $^{+800 }  _{-600  } $   &   46\,000  $^{+4\,000 }_{-3\,000 }  $  &    14.95 $ ^{+0.01 } _{-0.04 } $ &   0.282 $^{+0.009} _{-0.035}$ &   -3.58 $^{+0.21} _{-0.10}$  &  -4.89 $^{+0.28} _{-0.23}$  &  131 $^{+10}_{-48} $ & $221^{+10}_{-10} $ \\ \\
SMP LMC 66   &  4\,500  $^{+2\,000} _{-1\,500 } $  &   107\,000 $^{+5\,000 }_{-7\,000 }  $  &    14.76 $^ {+0.07 } _{-0.06 } $ &   0.205 $^{+0.041} _{-0.028}$ &   -3.13 $^{+0.12} _{-0.13}$  &  $-$                        &  102 $^{+38}_{-24} $ & $ - $   \\ \\
SMP LMC 71   &  5\,400  $^{+400 }  _{-100  } $   &   164\,000 $^{+9\,000 }_{-4\,000 }  $  &    14.15 $ ^{+0.01 } _{-0.02 } $ &   0.065 $^{+0.002} _{-0.005}$ &   -2.21 $^{+0.15} _{-0.10}$  &  -3.46 $^{+0.09} _{-0.19}$  &  114 $^{+23}_{-24} $ & $197^{+10}_{-10} $  \\ \\
SMP LMC 102  &  3\,600  $^{+900 }  _{-800  } $   &   140\,000 $^{+9\,000 }_{-10\,000}  $  &    15.29 $^ {+0.03 } _{-0.06 } $ &   0.370 $^{+0.042} _{-0.064}$ &   -3.52 $^{+0.12} _{-0.20}$  &  $-$                        &  90  $^{+22}_{-21} $ & $ -   $  \\ \\

\hline
  ORD &             &            &         &   &  log($\delta$[Sil])                           &   & $\rm{T}_d$[Sil] [K]     &  \\
\hline
\\
SMP LMC 81   &  4\,700  $^{+1\,200}  _{-800  } $   &   80\,000  $^{+40\,000}_{-15\,000}  $  &   $ 14.77  ^{+0.02 } _{-0.02 } $ &  $ 0.129^{+0.006} _{-0.009}$ & $-2.47^{+0.09}_{-0.10}$  &  $- $  & $ 102 ^{+22}_{-20}  $         & $-$        \\ \\
\hline   
Dust free
\\
\hline
\\
SMP LMC 80   &   3\,200  $^{+900 }  _{-1\,100 } $   &   57\,000 $ ^{+5\,000 }_{-3\,000 }  $  &   $ 14.45  ^{+0.04 } _{-0.04 } $ &  $ 0.051^{+0.007} _{-0.005}$ & $-$  &  $- $  & $- $   &  $ -$               \\ \\
\hline
\label{tabpost}
\end{tabular}
\\ \textbf{Notes.} In the columns are reported: source ID; luminosity and effective temperature of the CS;  thickness and mass of the gas of the nebula; logarithmic dust-to-gas mass ratio of carbon or silicate dust and PAHs; carbon or silicate dust temperature; PAH temperature.
\end{table*}

\begin{figure*}
\vskip-40pt
\begin{minipage}{0.46\textwidth}
\resizebox{1.\hsize}{!}{\includegraphics{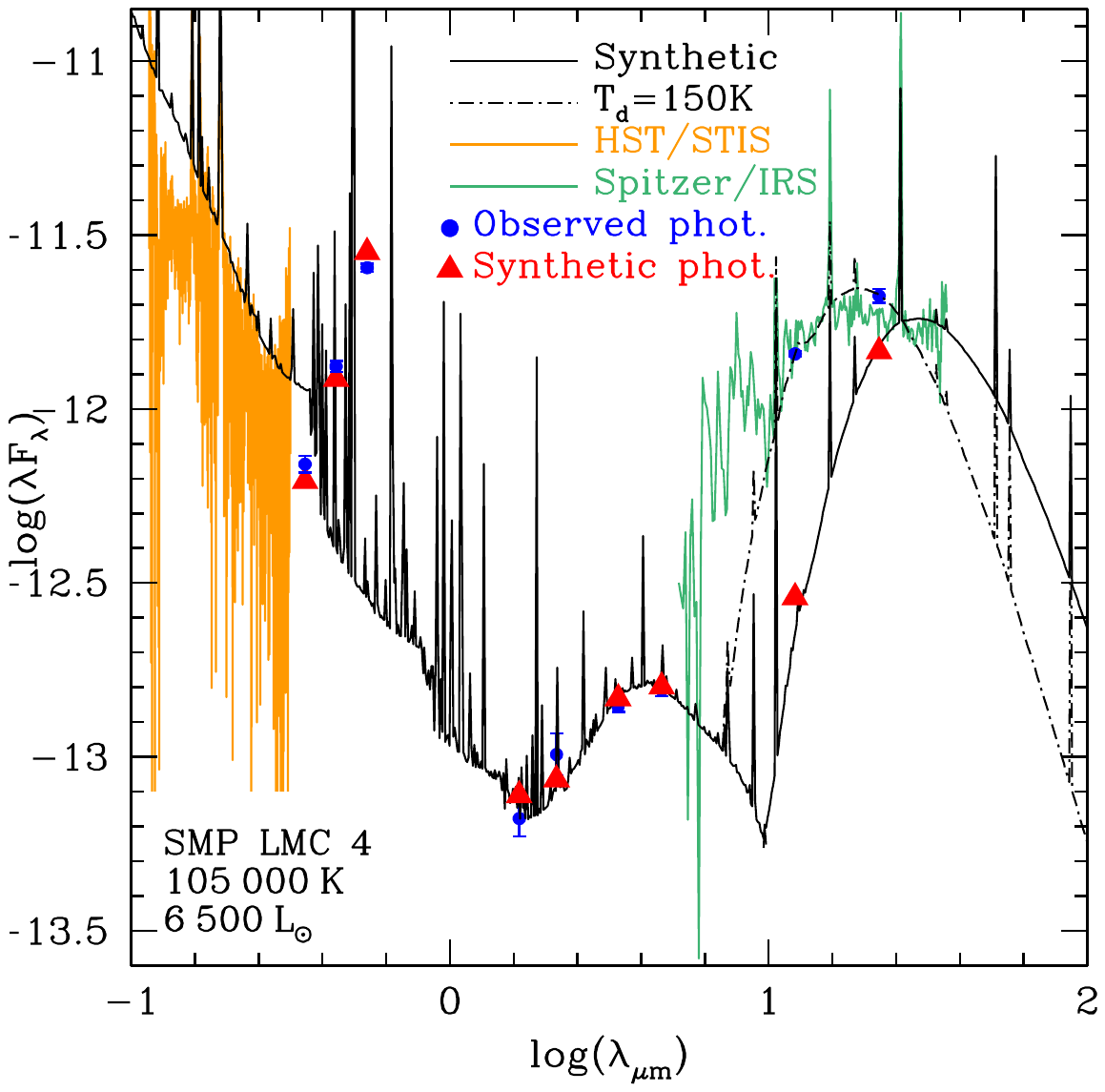}}
\end{minipage}
\begin{minipage}{0.46\textwidth}
\resizebox{1.\hsize}{!}{\includegraphics{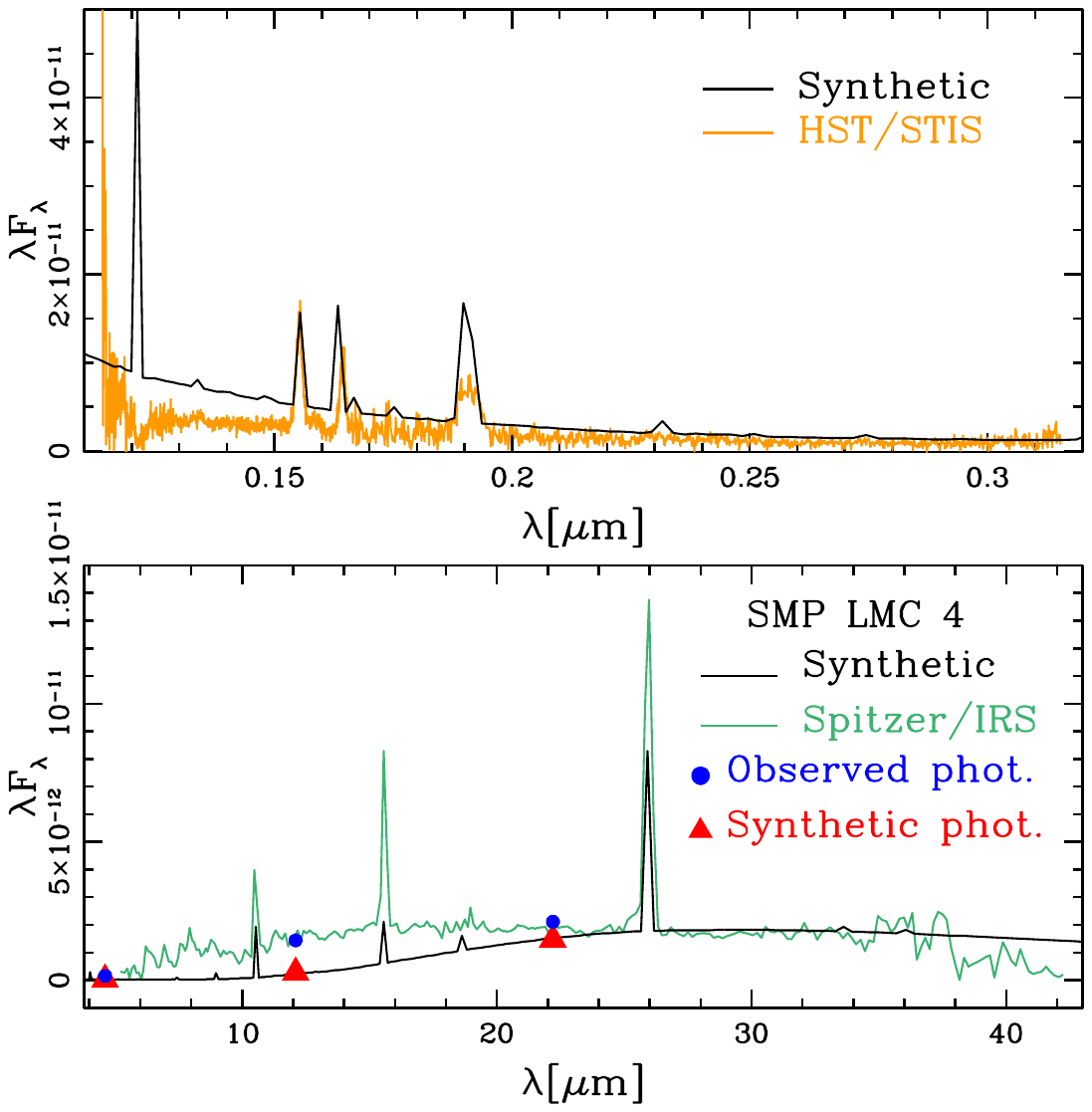}}
\end{minipage}
\vskip-90pt
\begin{minipage}{0.46\textwidth}
\resizebox{1.\hsize}{!}{\includegraphics{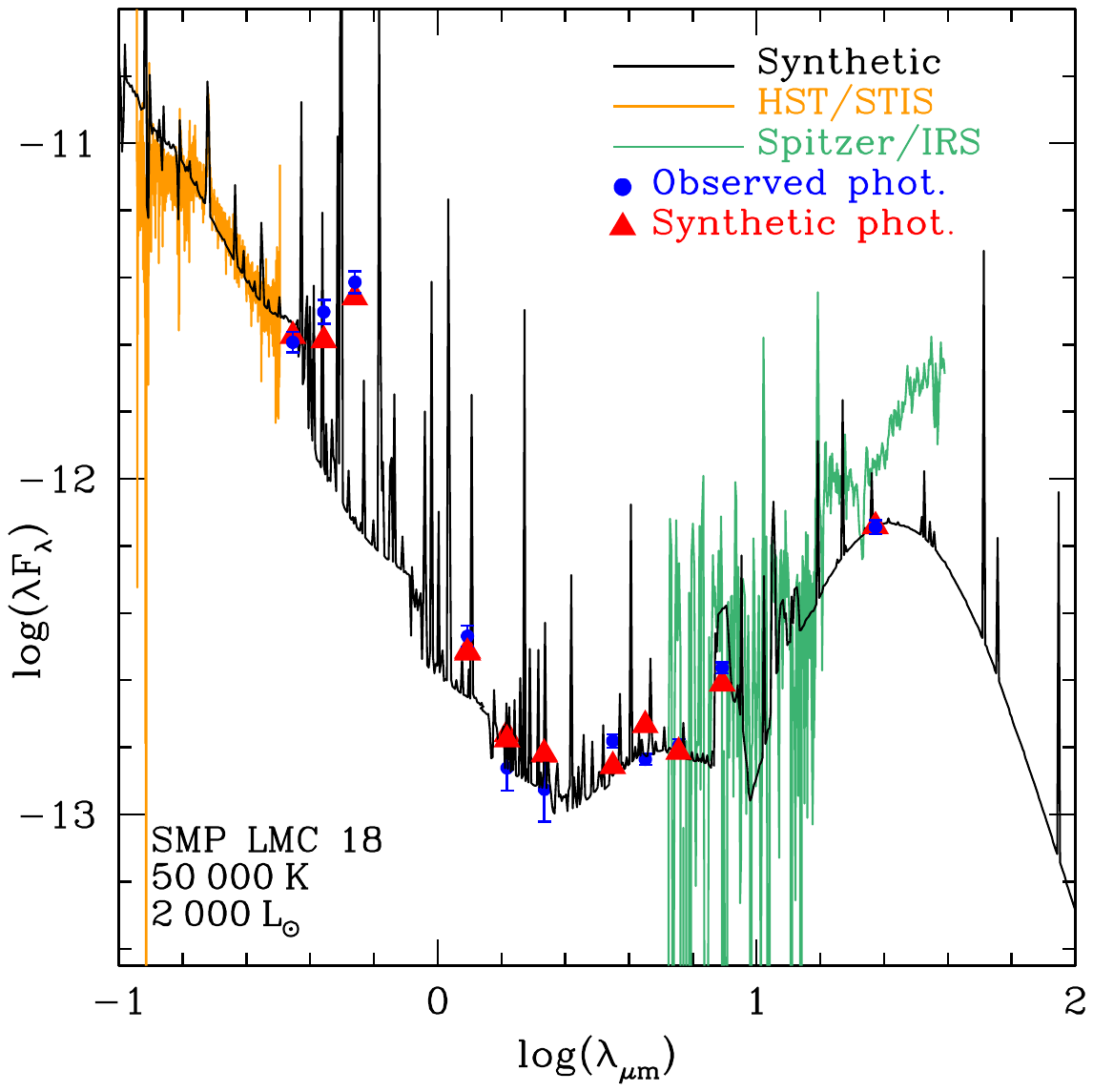}}
\end{minipage}
\begin{minipage}{0.46\textwidth}
\resizebox{1.\hsize}{!}{\includegraphics{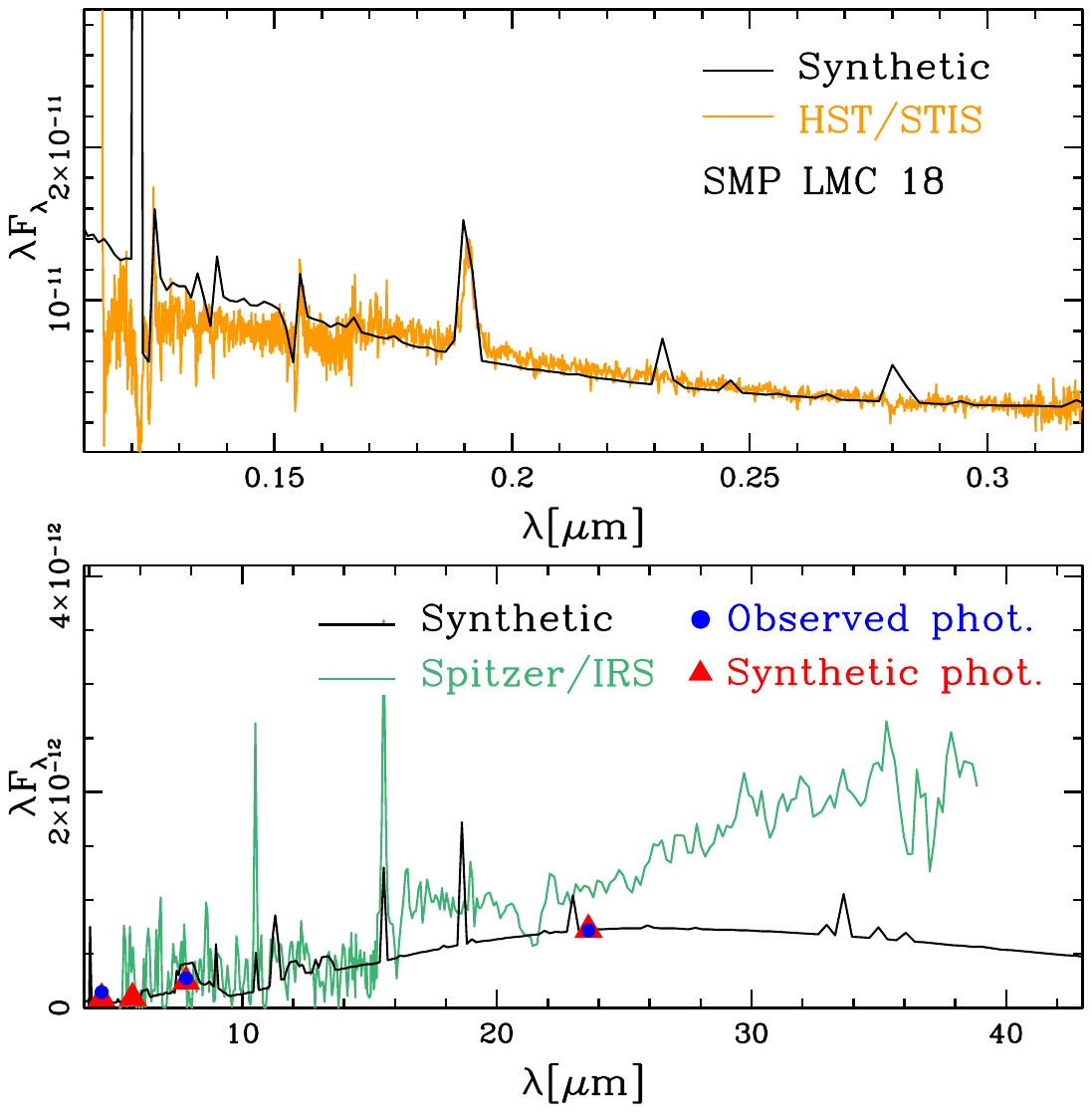}}
\end{minipage}
\vskip-90pt
\begin{minipage}{0.46\textwidth}
\resizebox{1.\hsize}{!}{\includegraphics{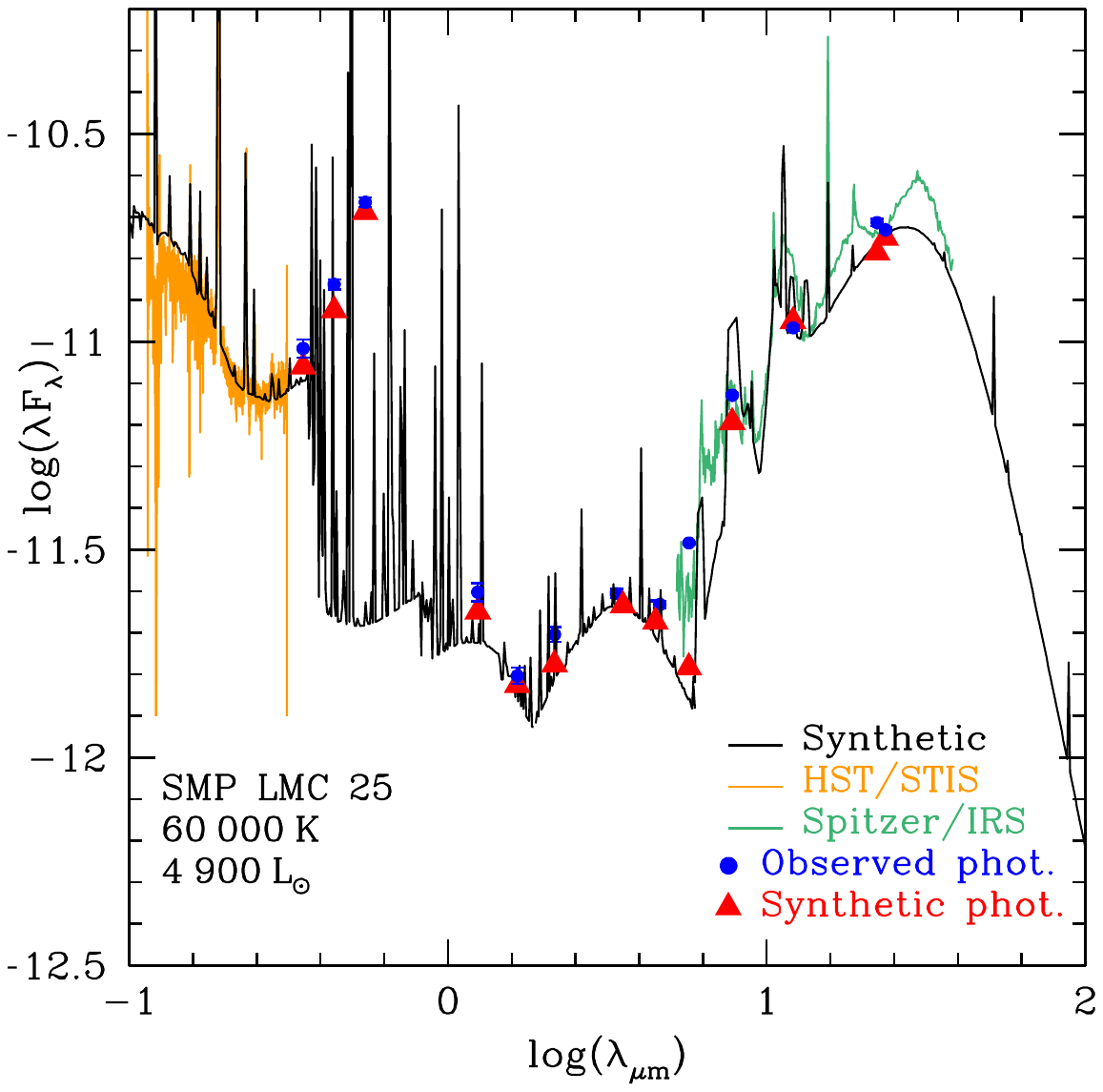}}
\end{minipage}
\begin{minipage}{0.46\textwidth}
\resizebox{1.\hsize}{!}{\includegraphics{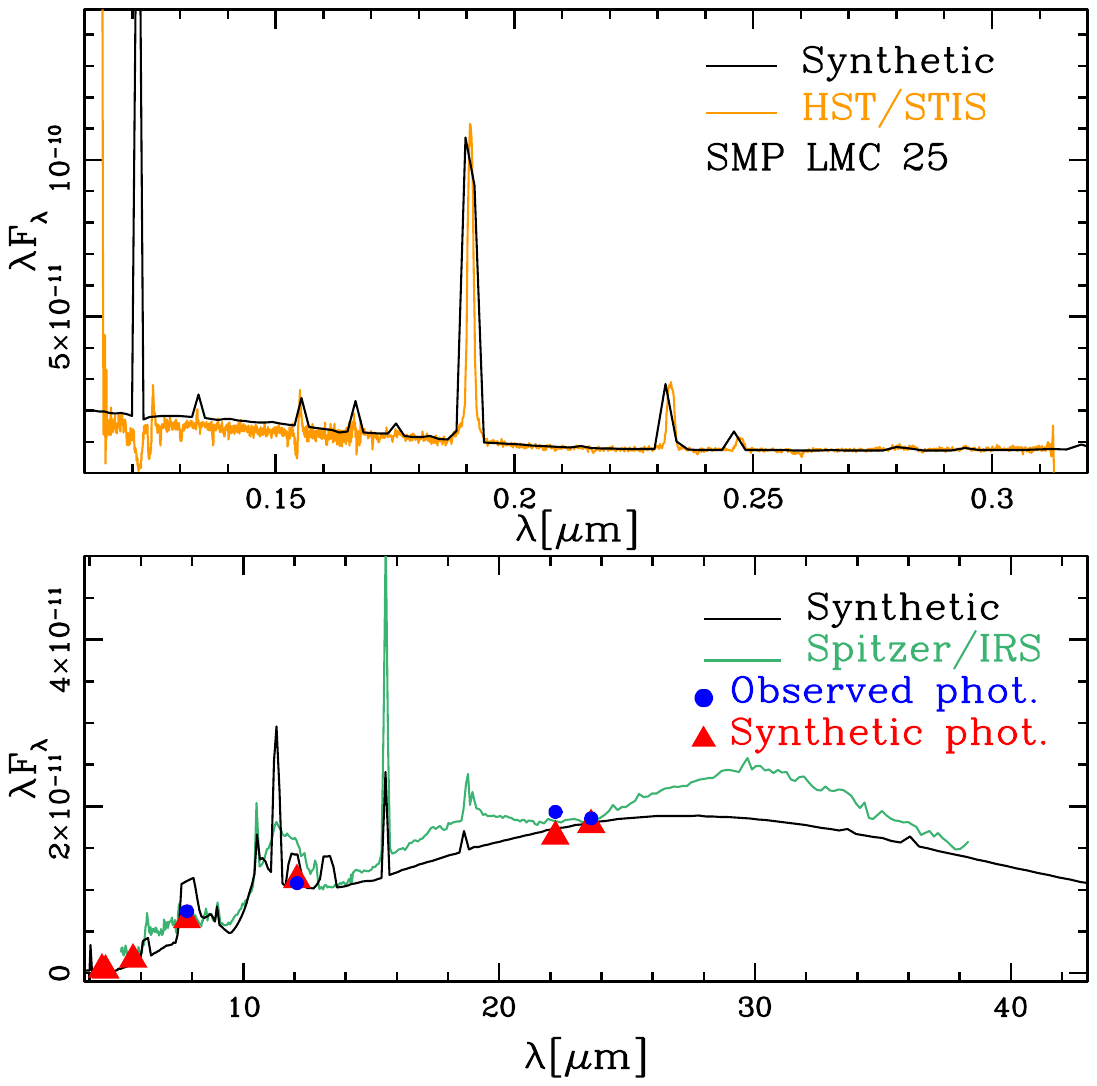}}
\end{minipage}
\vskip-60pt
\caption{SED of the CRD PNe, composed of the photometric data (blue dots) from \cite{reid14}, \cite{cutri12}, and \cite{lasker08}, the HST/STIS UV spectrum taken from \cite[in orange]{stanghellini05}, and the \textit{Spitzer}/IRS spectrum from \cite[in green]{stanghellini07}. The black lines and red triangles indicate the synthetic spectra and photometry obtained in this work. The dotted line in the upper-left panel is the synthetic spectrum obtained by assuming a dust temperature of 150\,K. In the right panels there are zoomed-in views of the UV (upper panel) and the IR spectra (lower panel). 
} 
\label{f1}
\end{figure*}

\setcounter{figure}{1}
\begin{figure*}
\vskip-40pt
\begin{minipage}{0.46\textwidth}
\resizebox{1.\hsize}{!}{\includegraphics{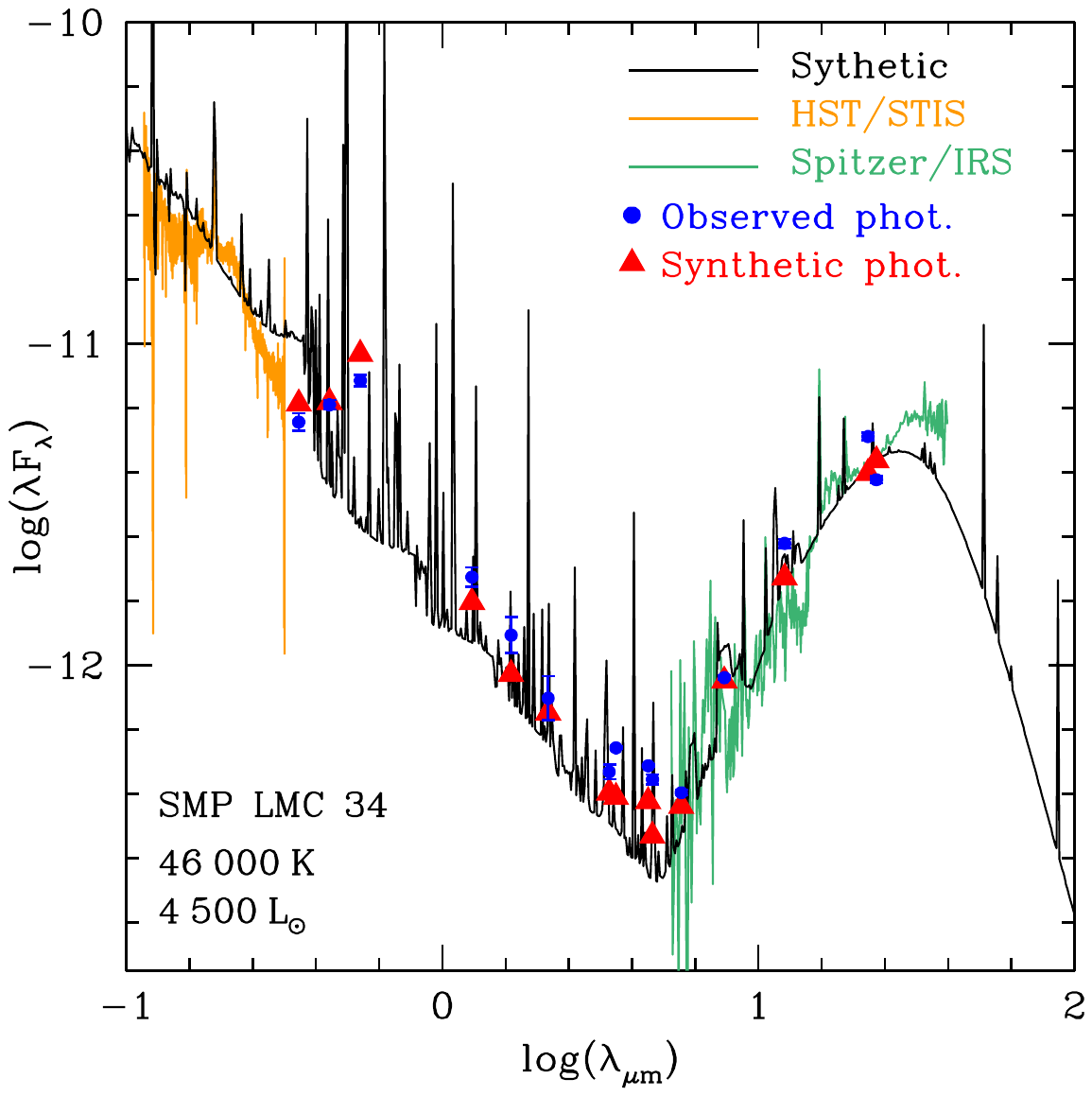}}
\end{minipage}
\begin{minipage}{0.46\textwidth}
\resizebox{1.\hsize}{!}{\includegraphics{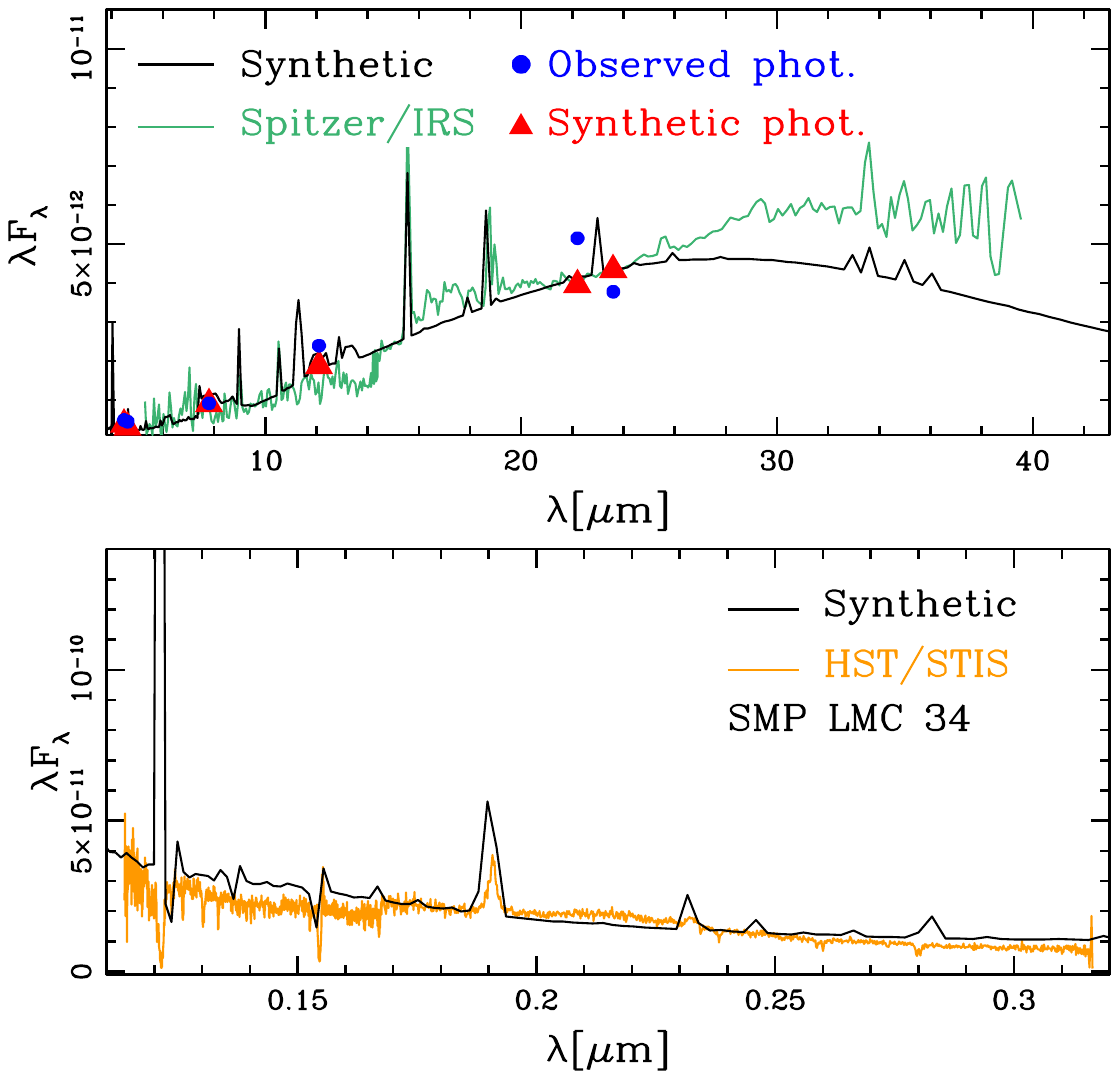}}
\end{minipage}
\vskip-90pt
\begin{minipage}{0.46\textwidth}
\resizebox{1.\hsize}{!}{\includegraphics{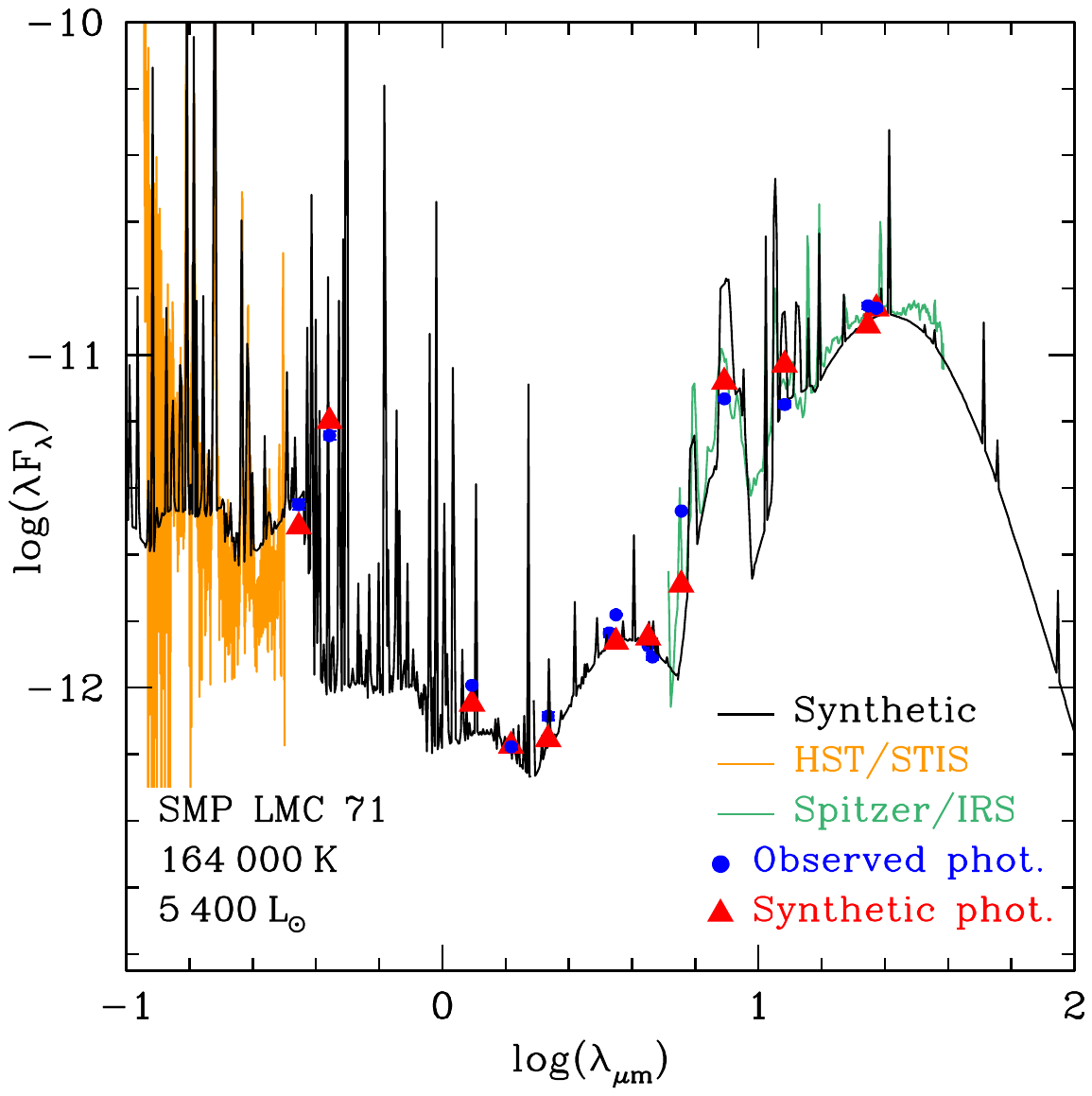}}
\end{minipage}
\begin{minipage}{0.46\textwidth}
\resizebox{1.\hsize}{!}{\includegraphics{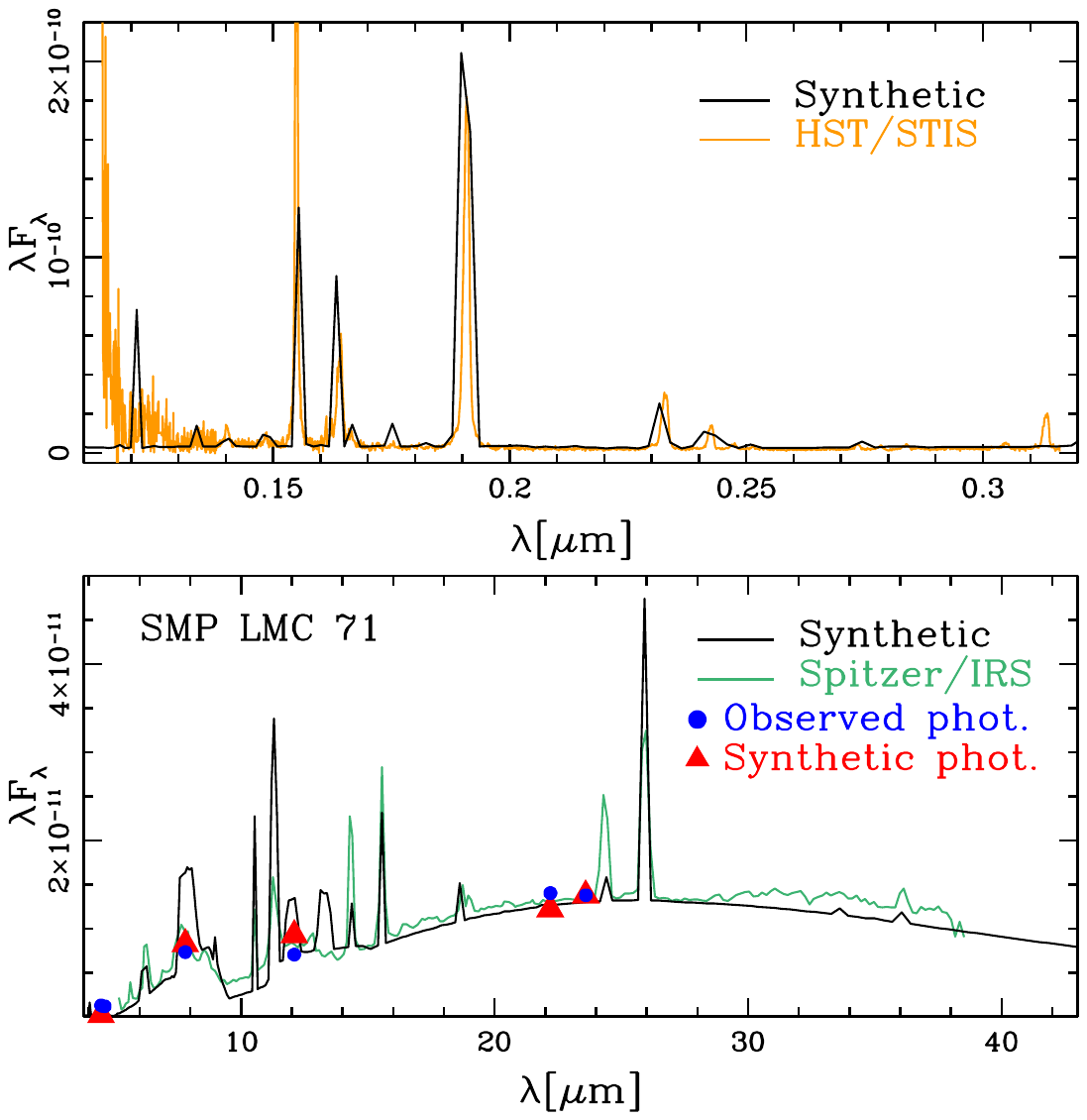}}
\end{minipage}
\vskip-90pt
\begin{minipage}{0.46\textwidth}
\resizebox{1.\hsize}{!}{\includegraphics{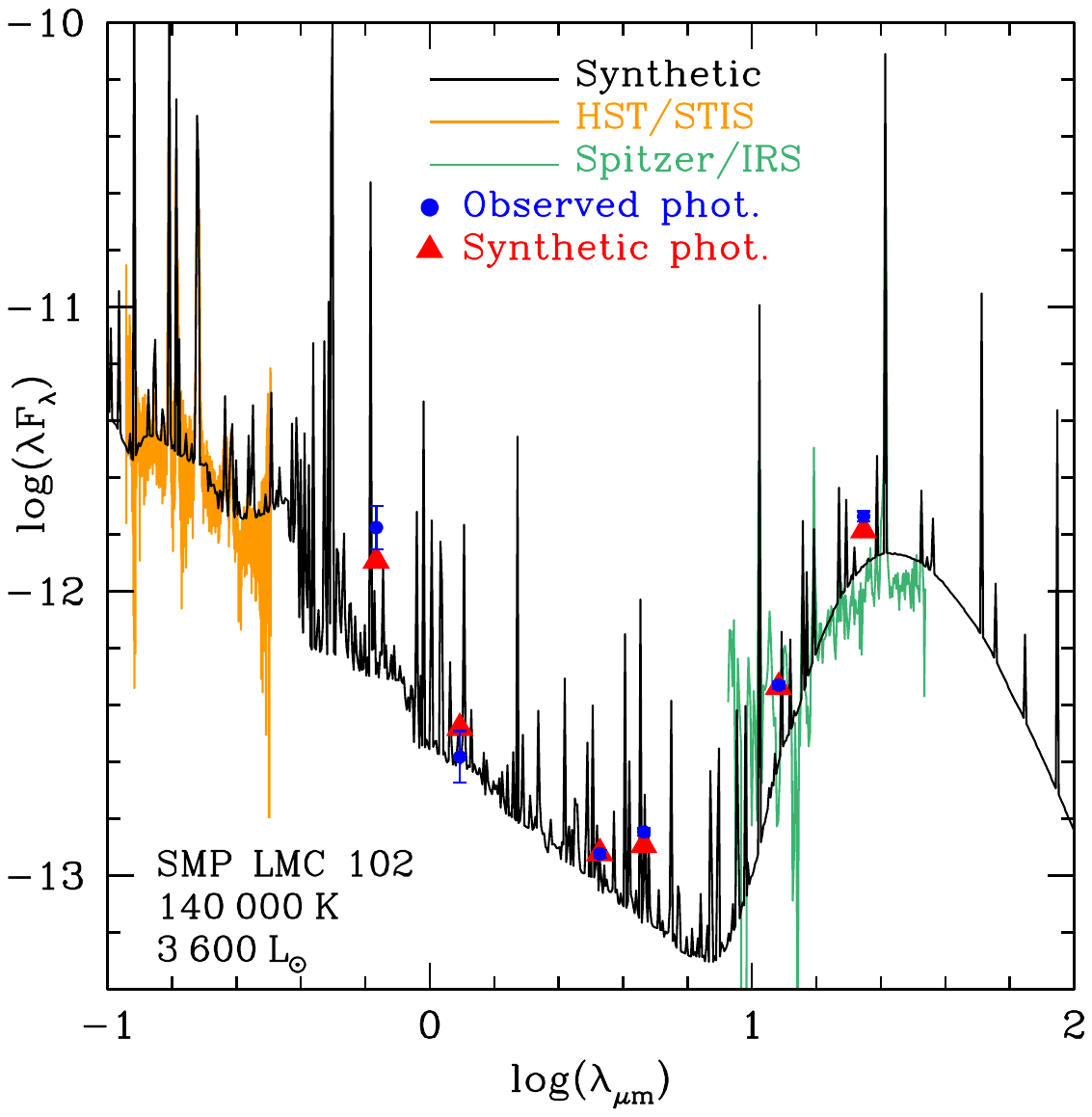}}
\end{minipage}
\begin{minipage}{0.46\textwidth}
\resizebox{1.\hsize}{!}{\includegraphics{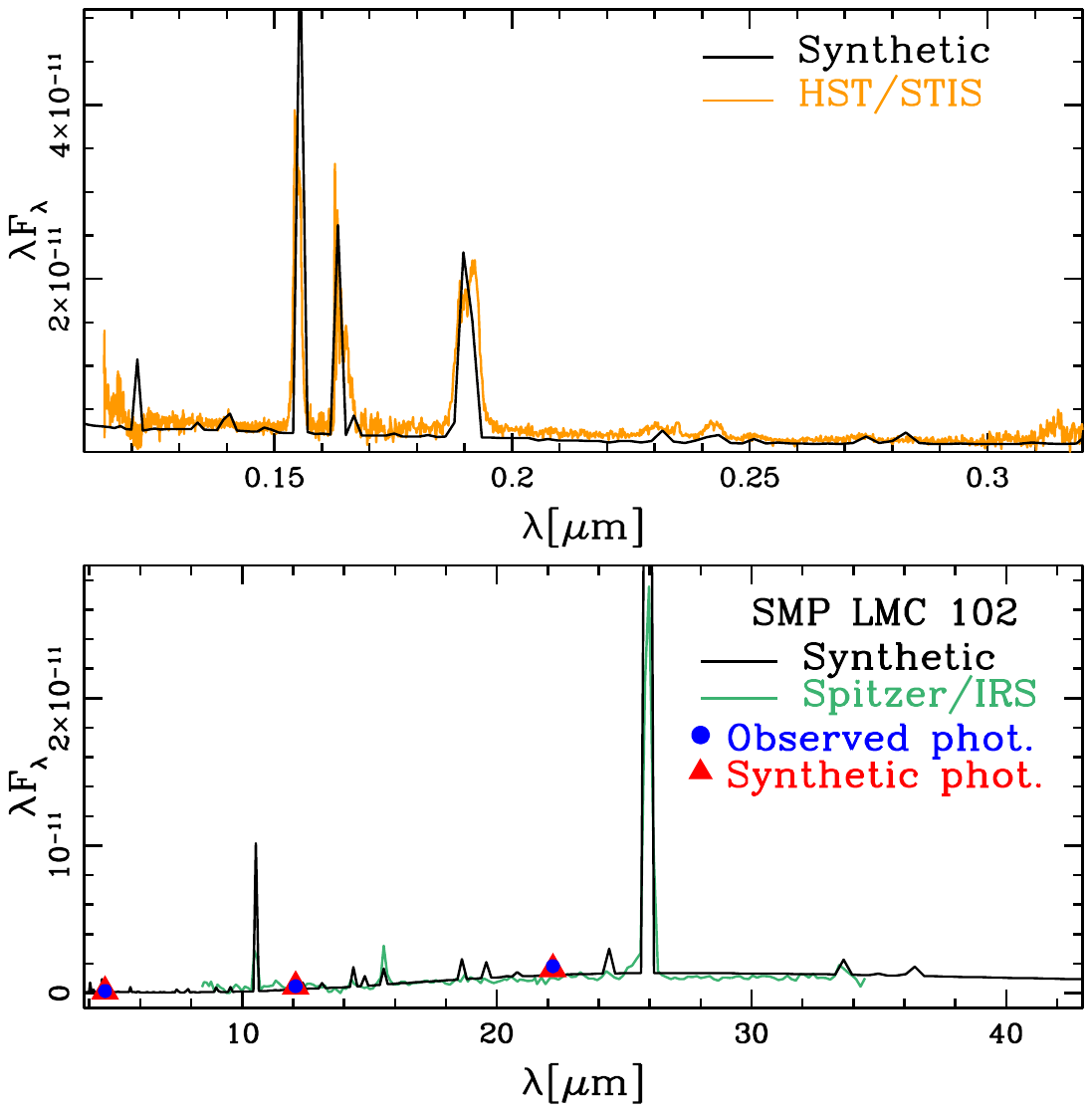}}
\end{minipage}
\vskip-60pt
\caption{ Continued.} 
\label{f2}
\end{figure*}

\begin{figure*}
\vskip-40pt
\begin{minipage}{0.46\textwidth}
\resizebox{1.\hsize}{!}{\includegraphics{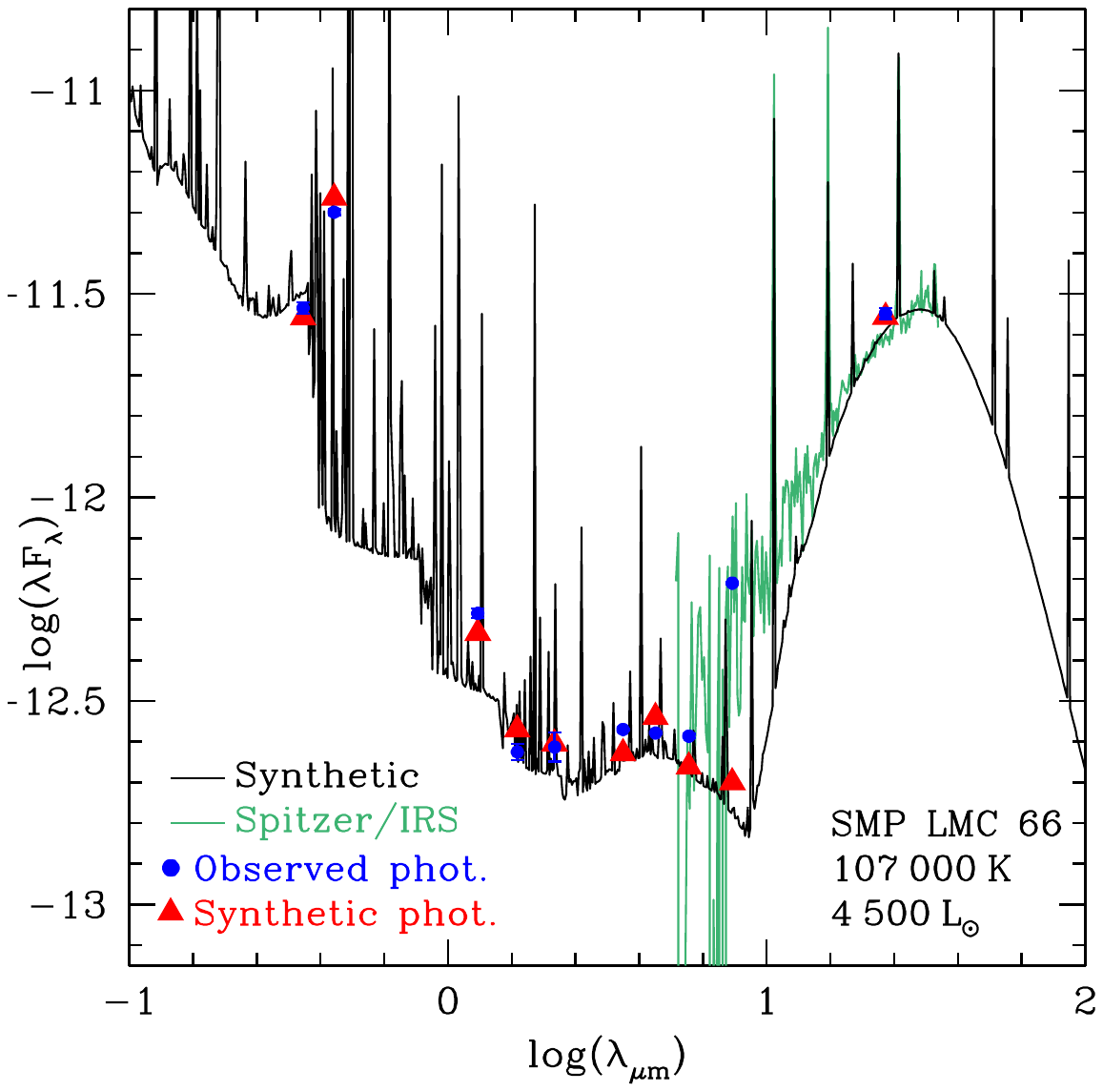}}
\end{minipage}
\begin{minipage}{0.46\textwidth}
\resizebox{1.\hsize}{!}{\includegraphics{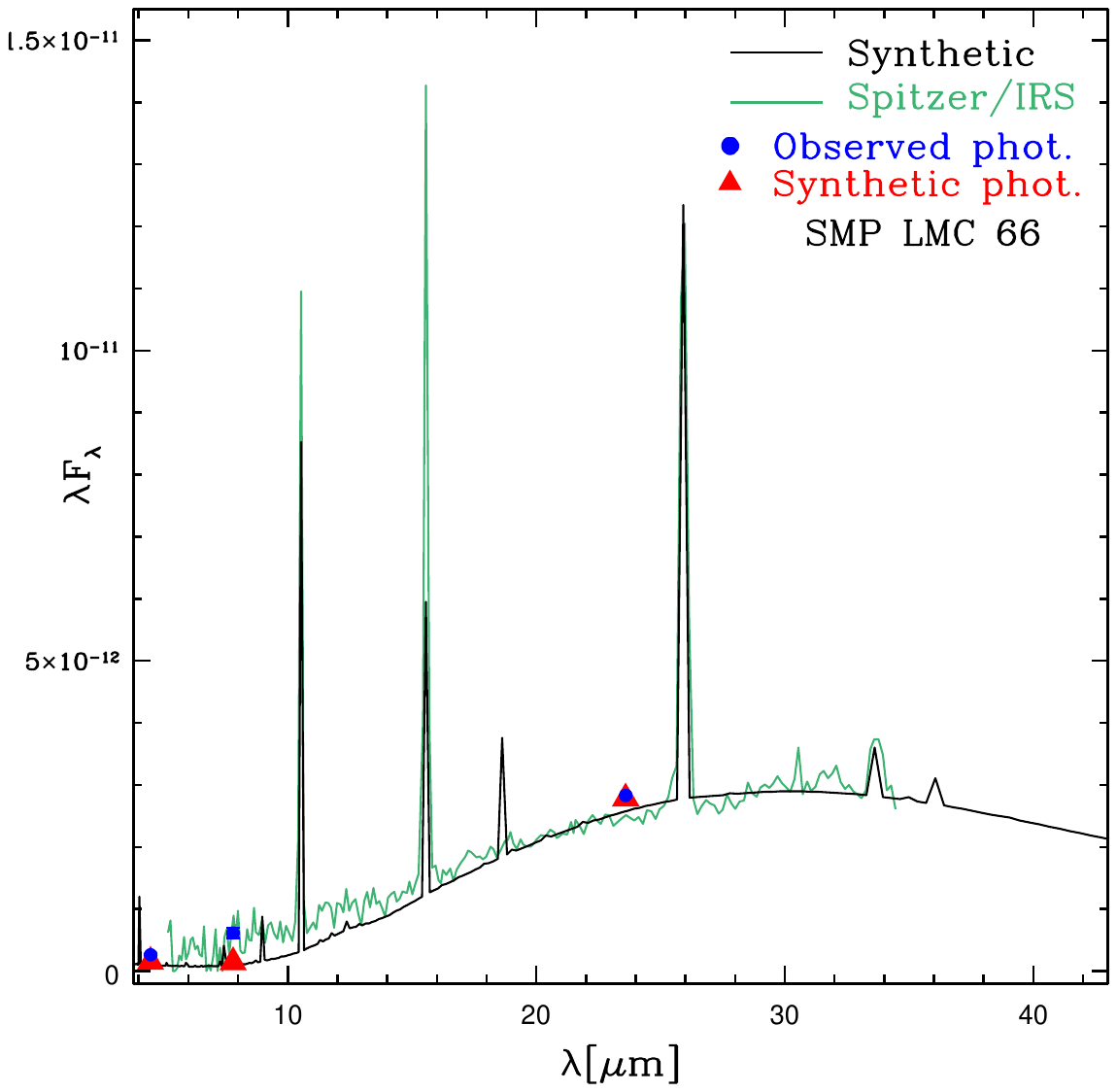}}
\end{minipage}
\vskip-60pt
\caption{SED of the CRD PN SMP LMC 66, composed of the photometric data (blue dots) from \cite{reid14}, \cite{cutri12}, and \cite{lasker08} and the \textit{Spitzer}/IRS spectrum from \cite[in green]{stanghellini07}. The black line and red triangles indicate the synthetic spectrum and photometry obtained in this work. In the right panel there is a zoomed-in view of the IR spectrum. 
} 
\label{f3}
\end{figure*}

\begin{figure*}
\vskip-40pt
\begin{minipage}{0.46\textwidth}
\resizebox{1.\hsize}{!}{\includegraphics{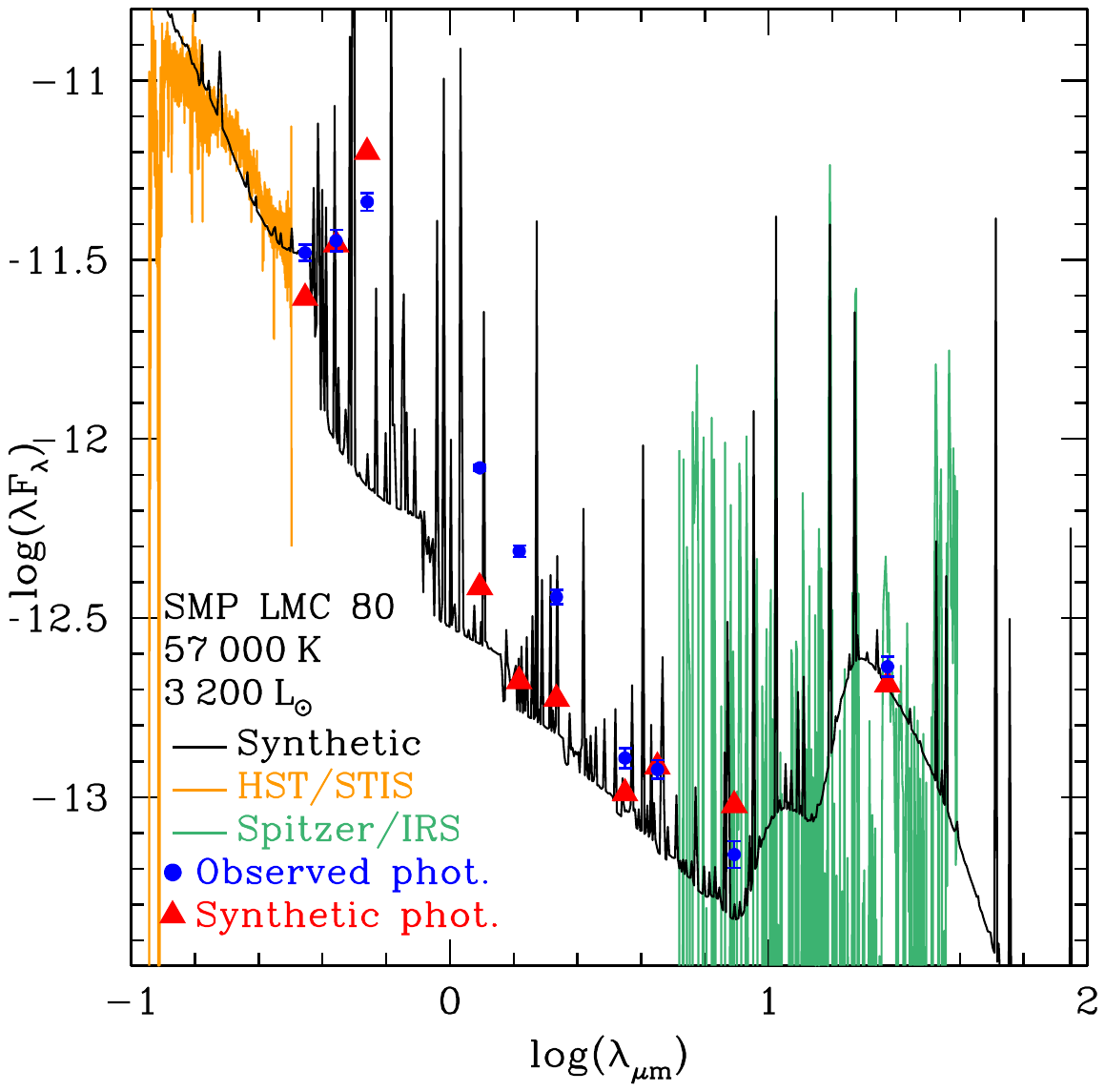}}
\end{minipage}
\begin{minipage}{0.46\textwidth}
\resizebox{1.\hsize}{!}{\includegraphics{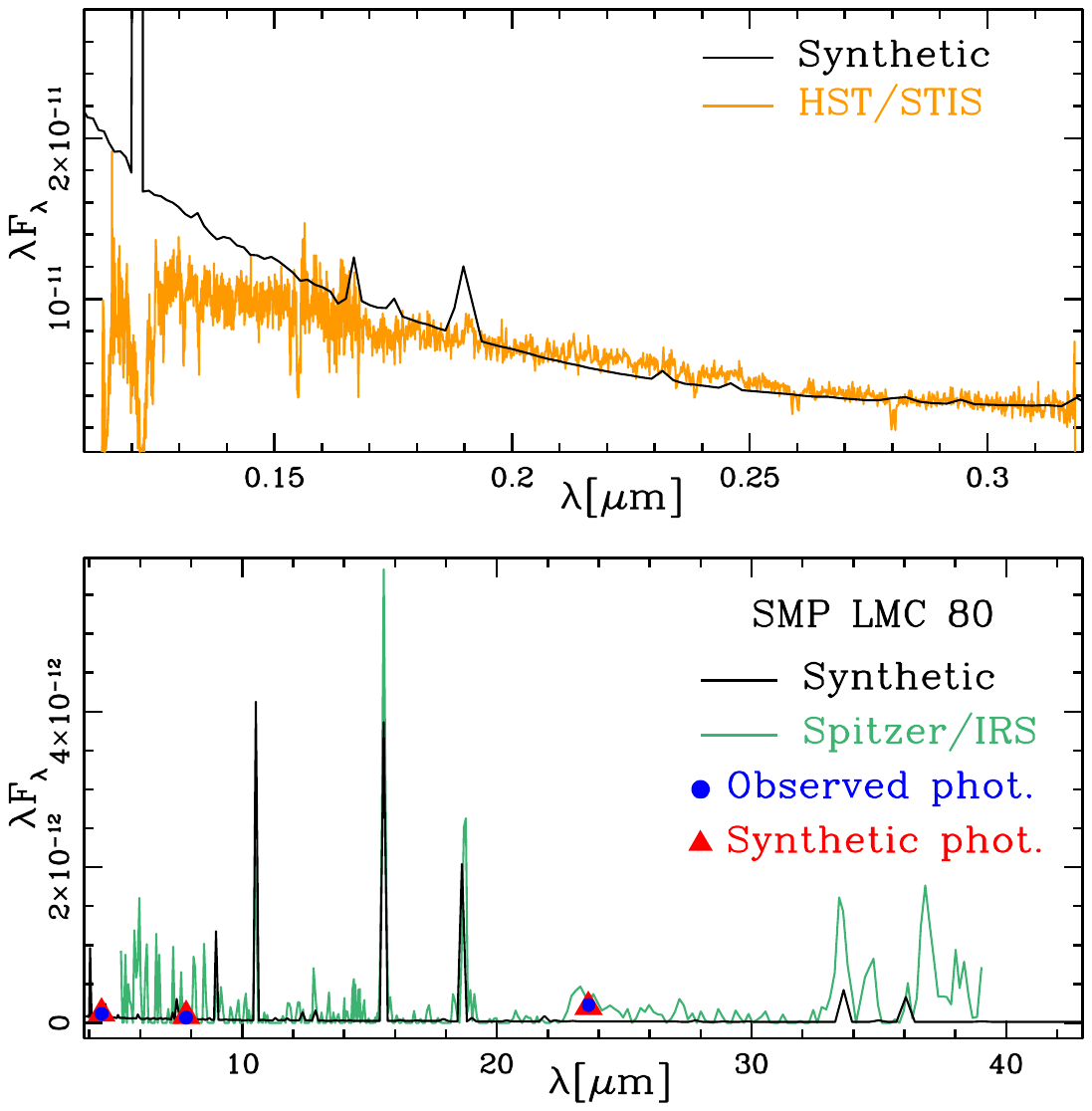}}
\end{minipage}
\vskip-90pt
\begin{minipage}{0.46\textwidth}
\resizebox{1.\hsize}{!}{\includegraphics{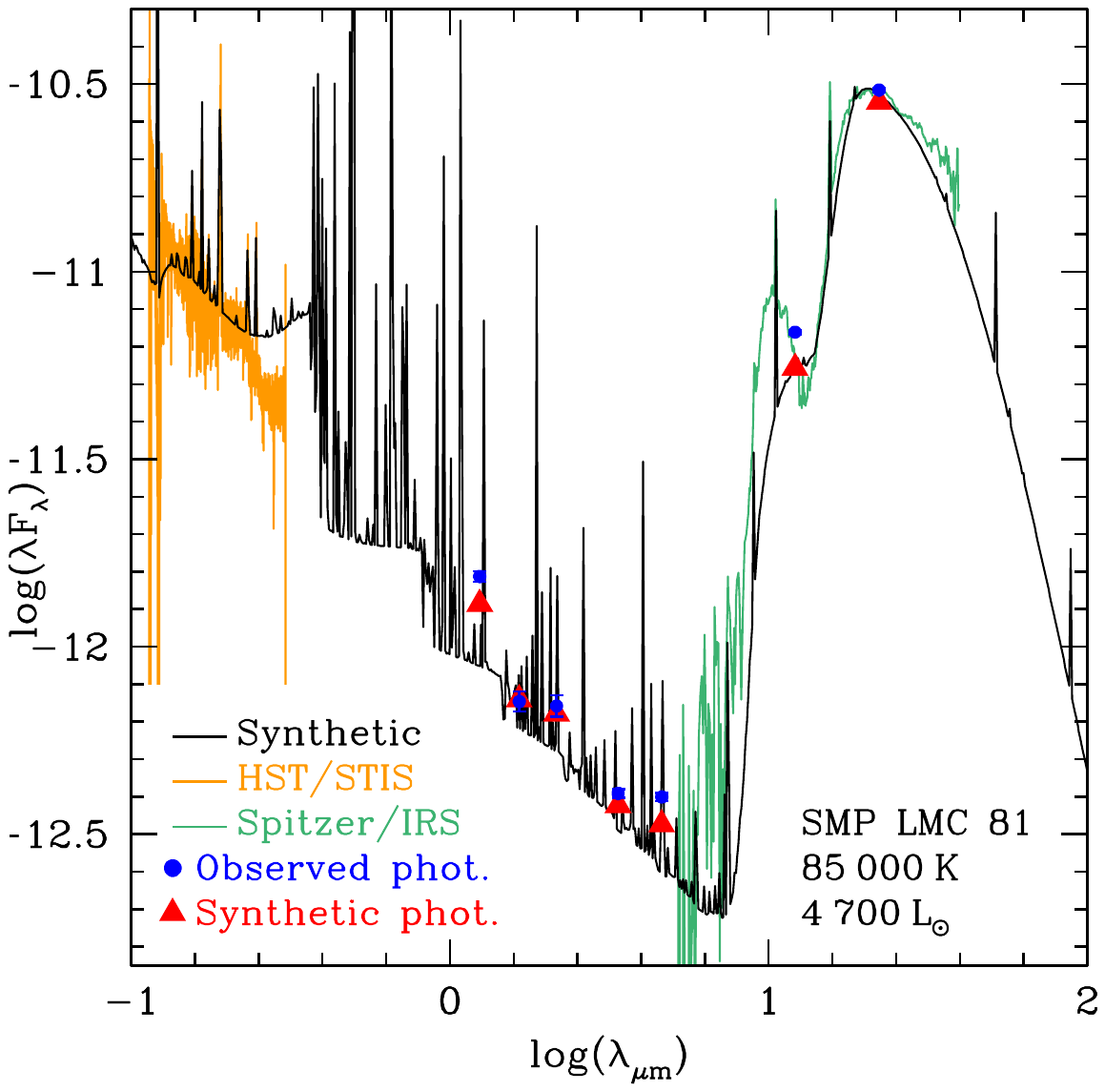}}
\end{minipage}
\begin{minipage}{0.46\textwidth}
\resizebox{1.\hsize}{!}{\includegraphics{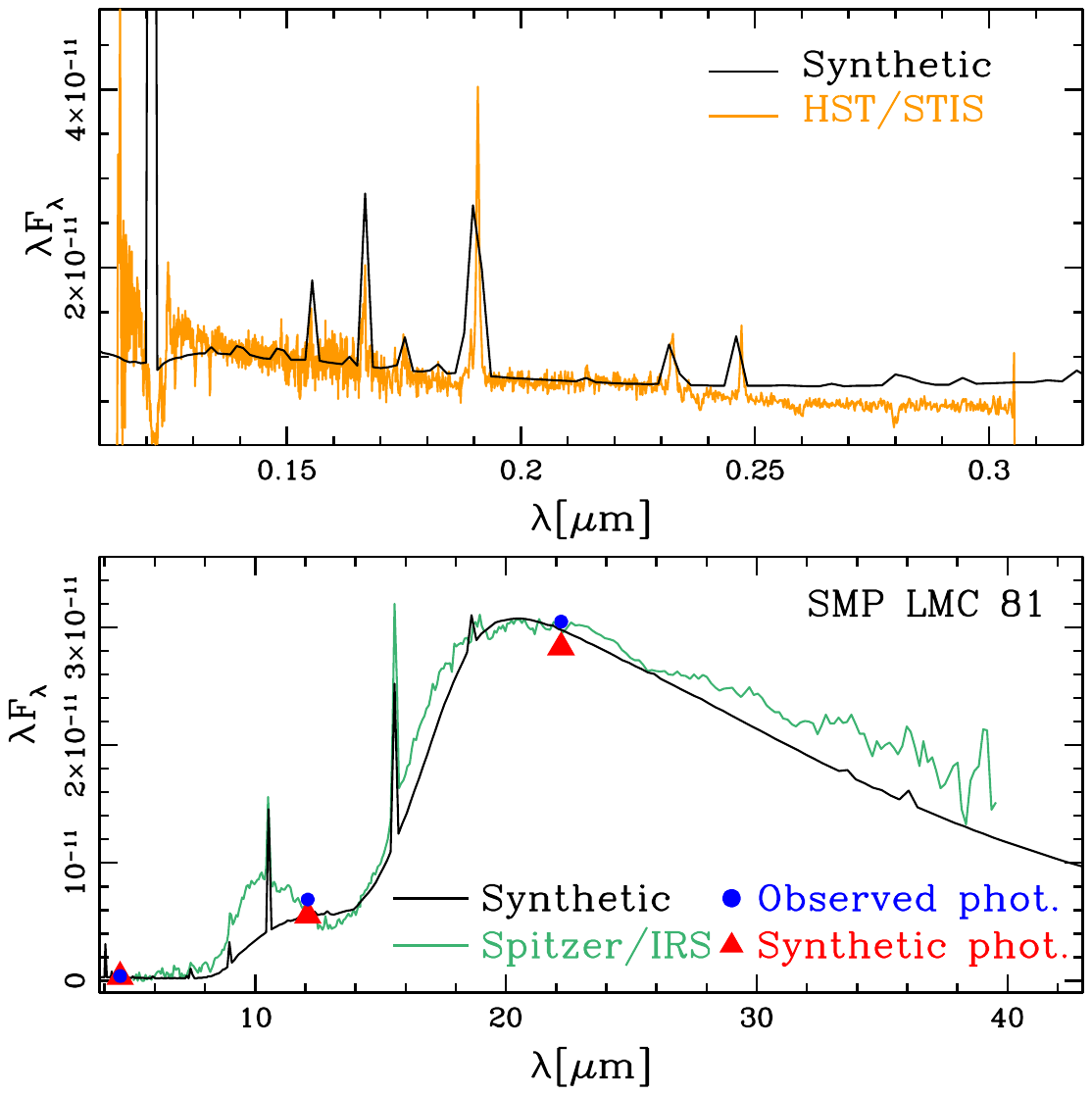}}
\end{minipage}
\vskip-60pt
\caption{
SEDs of the ORD PNe, composed of the photometric data (blue dots) from \cite{reid14}, \cite{cutri12}, and \cite{lasker08}, the HST/STIS UV spectrum taken from \cite[in orange]{stanghellini05}, and the \textit{Spitzer}/IRS spectrum from \cite[in green]{stanghellini07}. The black line and red triangles indicate the synthetic spectra and photometry obtained in this work. In the right panels there are zoomed-in views of the UV (upper panel) and the IR spectra (lower panel).  
} 
\label{f4}
\end{figure*}

\section{SED analysis of the target sample}
\label{SED}

From the comparison of the observational data (i.e., abundances, spectra, and photometry, described in Sect. \ref{obs}) and the photoionization model output (see Sect. \ref{input}), we were able to determine the temperature and luminosity of the CSs, the nebular thickness, the nebular gaseous mass, the dust-to-gas ratio, and the dust temperature. These results are given in Table \ref{tab1}.

In Figures 2$-$5 we plot the synthetic spectrum produced using CLOUDY, the HST/STIS (LS05), and \textit{Spitzer}/IRS (LS07) spectra. We are able to reproduce the main SED components in most PNe, with the exceptions due to the presence of broad unidentified dust features such as those at 6--9, 9--13, 15--20, and 25--35$\,\mu$m and/or other species for which the optical constants cannot be included in our photoionization modeling (see Sect. \ref{input} for details). The detailed comparison between our results and the observational data is extensively discussed in Appendix \ref{appendix}.

As reported in Table \ref{tab1} we find that the CSs of our sample are characterized by luminosities that span over 2\,000--6\,500\,$\rm{L}_\odot$ and by effective temperatures of 46\,000-164\,000\,K. We compared our temperatures with those obtained by EV03 adopting the Zanstra method \citep{zanstra1931}; a detailed discussion on this regard is reported in Appendix \ref{appendix}. The different morphology and structure of the nebulae reflects into a wide range of nebular thickness with $13.68\leq \rm{log}(\Delta\rm{R}/\rm{m})\leq 15.29$. As a result, the mass of the gas ranges from $0.034\,\rm{M}_\odot$ to $0.370\,\rm{M}_\odot$. $\rm{M}_{\rm{gas}}$ is a key parameter to understanding the connection of the PN stage with the previous ones and will be largely discussed in the second paper that is in preparation.

We find that the IR excess can be modeled in terms of carbonaceous dust in all but two PNe of our sample. The only exceptions are SMP LMC 80, which shows no evidence of dust emission, and SMP LMC 81, which shows a clear signature of silicate dust. For SMP LMC 34 we cannot rule out the possible presence of a small quantity of silicate dust in addition to the presence of carbonaceous dust. A detailed comparison of the modeling that we adopted and the results from the literature is reported in Sect. \ref{dust}. 

\section{Evolutionary history up to the PN phase: The ATON models}
\label{aton}

We aimed to reconstruct the evolutionary history of the targets from the AGB to the PN phase. To do so, we used evolutionary sequences computed using the stellar evolution code ATON \citep{ventura98}. 

The ATON models adopted in this study span from the pre-main sequence to the white dwarf cooling sequence. The code performs a full integration of the stellar structure equations, from the star's center to its photosphere, using the full spectrum of turbulence model \citep{cm91} to determine the temperature gradient within the convectively unstable regions. 
Associated with the evolutionary description of the CS, the ATON models follow the evolution of the dusty layer formed during the AGB phase, as described in \citet{ventura14}. An exhaustive description of the stellar evolution models, in particular those regarding the AGB phase and the related dust production, are presented in \cite{ventura22}. 

Recently, ATON computations have been extended to describe the evolution beyond the tip of the AGB. Details on the methodology adopted for the post-AGB and the following phases are given in \cite{devika23}. In the present work we further extend the library of models evolving to the PN phase, creating a wider set of evolutionary tracks that are more complete in terms of the considered mass and metallicity range. The models span the range from 0.6 to 2.5$\,\rm{M}_\odot$, with metallicities ranging from Z $=10^{-3}$ to $\rm{Z}=8\times 10^{-3}$. The masses considered in this context are expressed at the beginning of the AGB phase; this prevents any potential issues due to the still uncertain amount of mass lost during the red giant branch phase. This precaution is particularly relevant for the determination of the initial masses of stars with M $\leq 1.2\rm M_{\odot}$. We considered a wide range of metallicities, consistent with those known to contribute to the star formation history of the LMC \citep{harris09}. 

In Fig. \ref{tracks} we display the evolutionary sequences calculated with the stellar evolution code ATON in the Hertzsprung–Russell diagram, focusing  on the PN phase and varying the metallicity. As a general trend, we find that, for a fixed metallicity and effective temperature, higher luminosities correspond to higher masses. However, this trend does not hold for sources undergoing late thermal pulses, where late ignition of the helium shell causes the CS to evolve at lower luminosities compared to standard post-AGB evolution. Among the presented models, the $0.9\,\rm{M}_\odot$ model at $\rm{Z}=2\times 10^{-3}$ experiences this mechanism when it reaches effective temperatures of $\sim$15\,000\,K (see the right panel of Fig. \ref{tracks}). A deviation from the correlation between progenitor mass and luminosity is also found for the model of $2.5\,\rm{M}_\odot$ and $\rm{Z}=4\times 10^{-3}$ (see the left panel of Fig. \ref{tracks}). In this case, the reason lies in the TDU episodes experienced by the star. In Fig. \ref{f250} we show the variation of the surface carbon abundance (left panel) and of the core mass (right panel) as a function of the current mass of the star for three different masses: 1.5, 2.0, and 2.5\,M$_{\odot}$. Following the evolution of the 2.5\,M$_{\odot}$ model presented in this figure, we note that, when the mass of the star is $\sim$2\,M$_{\odot}$, the convective envelope depth during the TDU episode is such that the mass of the core is significantly reduced, becoming even lower than the core mass of the 2\,M$_{\odot}$ model. This phenomenon causes the 2.5\,M$_{\odot}$ star to evolve at a lower luminosity than its lower-mass counterpart, namely the 2\,M$_{\odot}$ model. The surface carbon and oxygen abundances, as computed at the end of the AGB phase, are reported in Table \ref{tabatonab} for all the models presented in this work.

%we can appreciate how the depth of the convective envelope during the TDU episode occurred when the mass of the star is $\sim$2\, M$_{\odot}$ is such that the mass of the core is significantly reduced, even lower than of the core mass of the 2\, M$_{\odot}$ model. This episode leads the star to evolve at a luminosity lower than the one of the lower mass counterpart, 2\, M$_{\odot}$ model. 

\section{Understanding the AGB-to-PN evolution of the target sample }
\label{mass}
\begin{figure*}
\vskip-40pt
\begin{minipage}{0.51\textwidth}
\resizebox{1.\hsize}{!}{\includegraphics{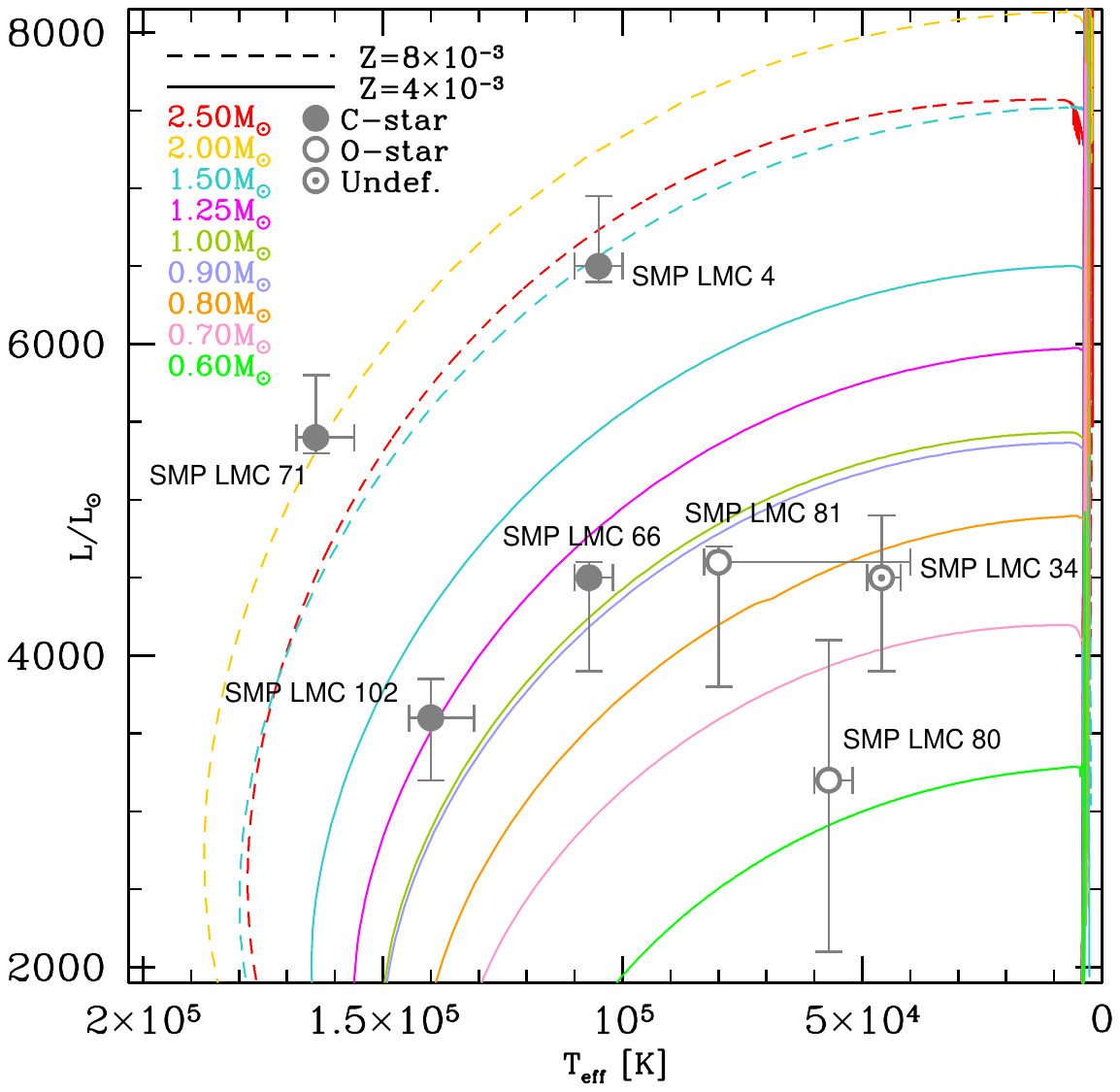}}
\end{minipage}
\begin{minipage}{0.51\textwidth}
\resizebox{1.\hsize}{!}{\includegraphics{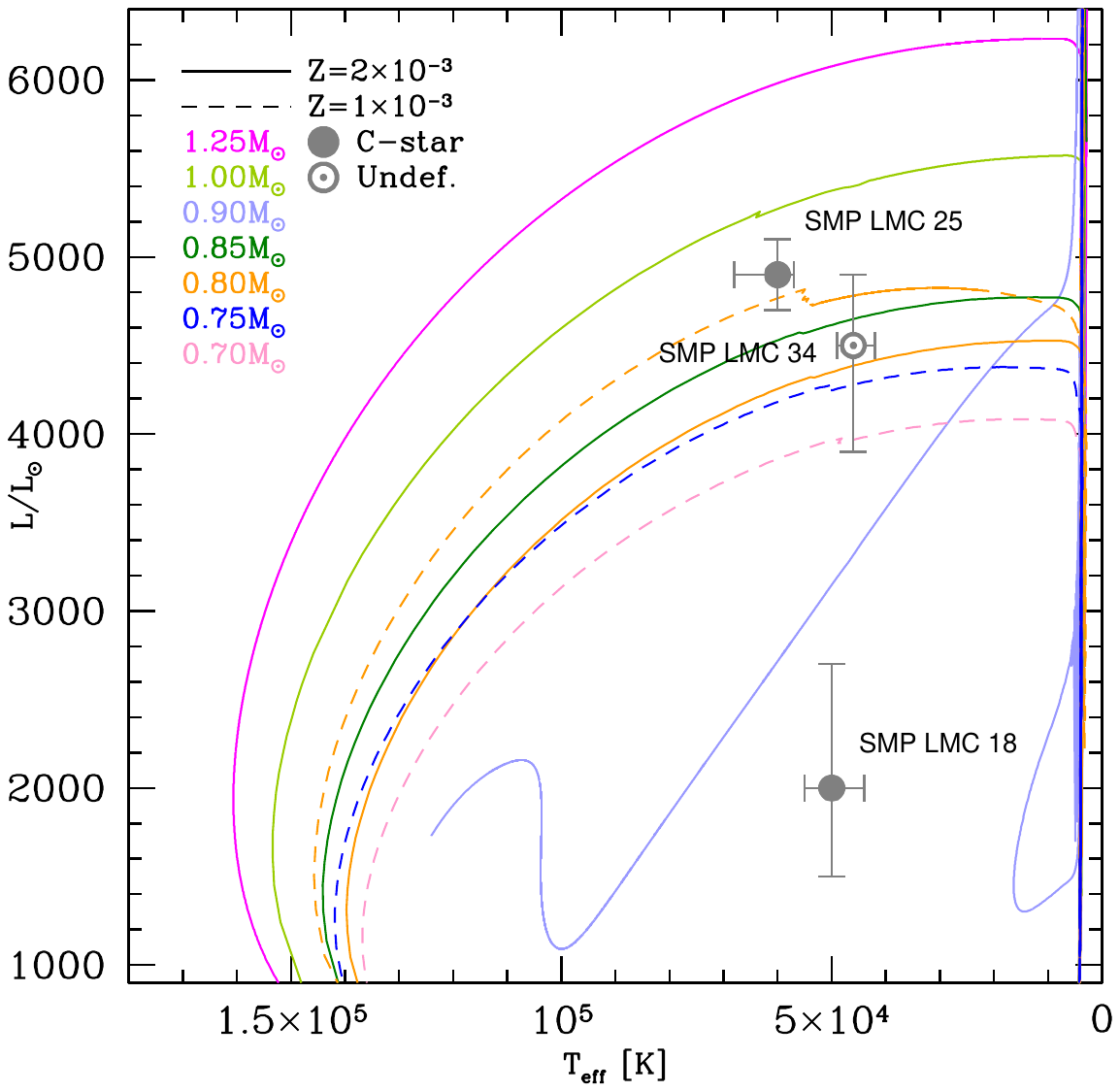}}
\end{minipage}
\vskip-60pt
\caption{Evolutionary tracks calculated with the stellar evolution code ATON \citep{ventura98} for different masses (different colors), expressed at the beginning of the AGB phase. The solid tracks are the $\rm{Z}=4 \times 10^{-3}$ (left panel) and  $\rm{Z}=2 \times 10^{-3}$ (right panel) and the dashed lines the $\rm{Z}=8 \times 10^{-3}$ (left panel) and $\rm{Z}=1\times 10^{-3}$ (right panel). The gray circles are the PN sample, with the luminosities and the effective temperatures derived in the present paper. Full and open symbols refer to C/O ratios greater and lower than unity, respectively. Due to the large oxygen error bar, the C/O of SMP LMC 34 is undefined.} 
\label{tracks}
\end{figure*}

\begin{figure*}
\vskip-40pt
\begin{minipage}{0.51\textwidth}
\resizebox{1.\hsize}{!}{\includegraphics{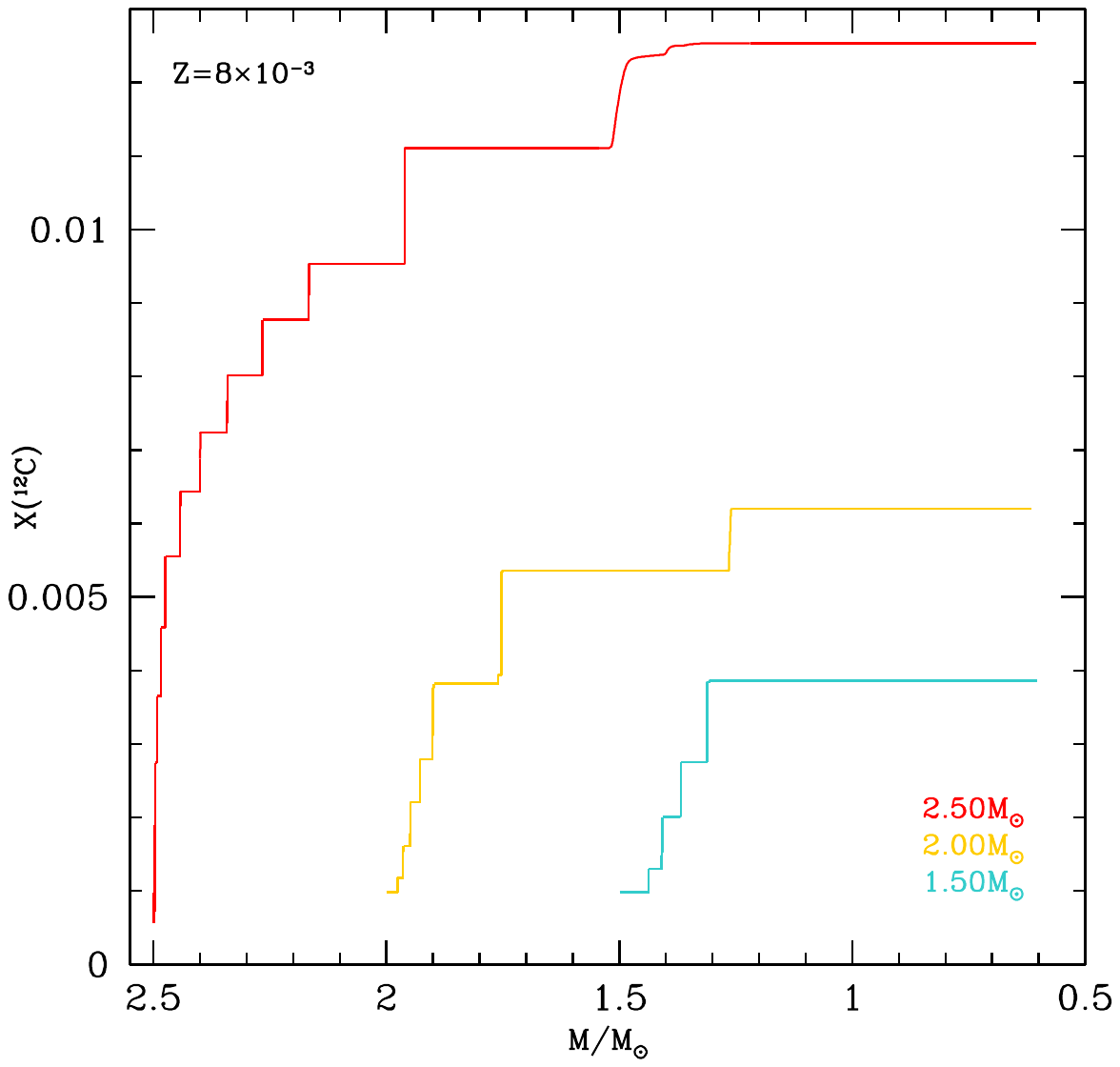}}
\end{minipage}
\begin{minipage}{0.51\textwidth}
\resizebox{1.\hsize}{!}{\includegraphics{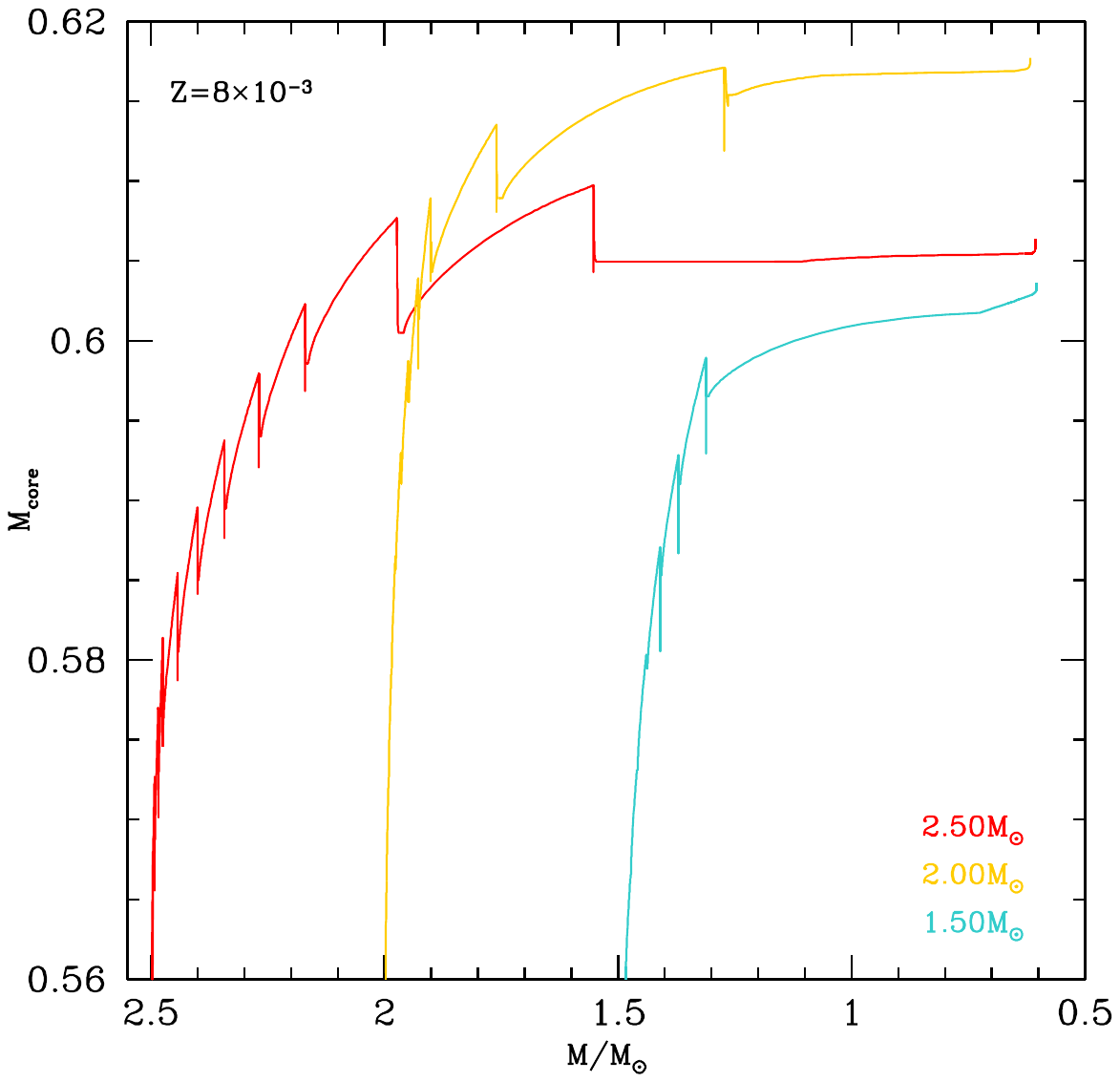}}
\end{minipage}
\vskip-60pt
\caption{Evolution of stellar parameters as a function of the of the actual mass of the CS, for three models at $\rm{Z}=8\times 10^{-3}$. \textbf{Left}: Surface carbon mass fraction for the $2.5\,\rm{M}_\odot$ model (red), $2.0\,\rm{M}_\odot$ (yellow), and $1.5~\rm{M}_\odot$ (cyan). \textbf{Right}: Core mass variation for the same models reported in the left panel.} 
\label{f250}
\end{figure*}

\begin{figure}
\vskip-40pt
\centering
\begin{minipage}{0.55\textwidth}
\addtolength{\leftskip}{-0.5cm}
\resizebox{1.\hsize}{!}{\includegraphics{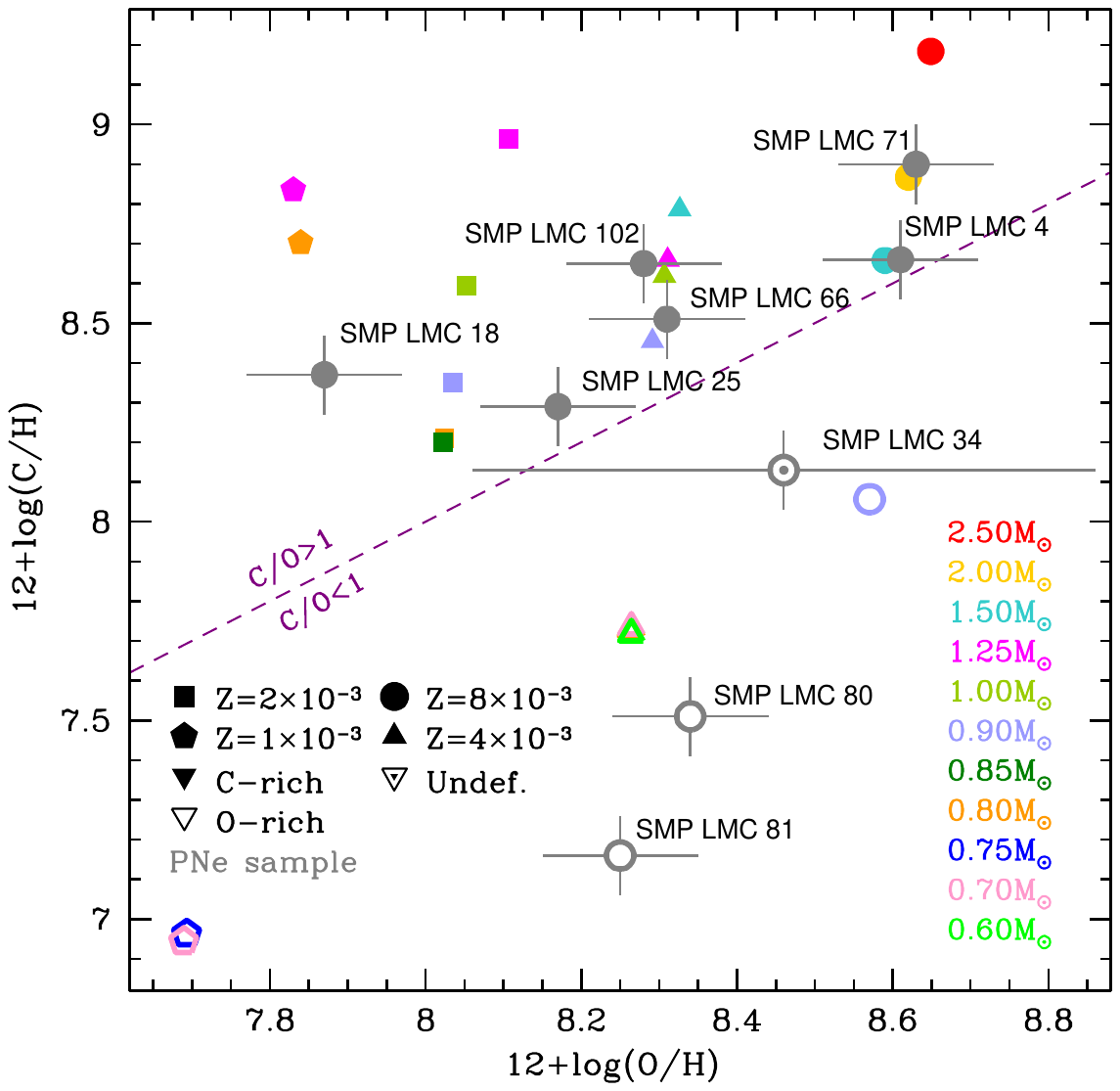}}
\end{minipage}
\vskip-60pt
\caption{Comparison of the abundances of carbon and oxygen observed for the PN sample (gray circles) and those computed using the ATON code at the end of the AGB evolution (colored symbols). Different shapes of the model represent different metallicities. Full and open symbols refer to the C/O ratio greater and lower than unity, respectively. The C/O of SMP LMC 34 is undefined due to the large oxygen error bar. The dashed line separates the oxygen-rich sources from the carbon-rich ones. } 
\label{fco}
\end{figure}

In the puzzling process of giving an evolutionary interpretation of the analyzed PNe, we took advantage of the evolutionary tracks presented in Fig. \ref{tracks} and their surface chemical abundances.  We compared the observed C and O abundance measurements (see Sect. \ref{obs}) with the surface chemical abundances of ATON models of different masses and metallicities (Table \ref{tabatonab}). This comparison is shown in Fig. \ref{fco}, where we present, in the C/H versus O/H plane, the comparison between the data (gray circles) and ATON results (colored symbols with different shapes according to the metallicity of the model). The ATON abundances are considered at the end of the AGB evolution, while the progenitor's masses are the ones at the beginning of the AGB phase. In Fig. \ref{fco} the dashed line indicates C/O=1; that is, carbon-rich sources are placed above this line. Due to the large oxygen error bar, the C/O of SMP LMC 34 is undefined (see Table \ref{tabinput}); it is represented with an open symbol with a central dot.

From the comparison of the data and model loci in the C/H versus O/H plane, we are able to infer the nature of the PN progenitor. In the mass regime $<$4\,M$_{\odot}$, the carbon abundance is a powerful tool for retrieving information on the progenitor's mass of post-AGB stars \citep{devika23} and PNe \citep{ventura15,letizia20,flavia23}. Stars with higher masses (2--3\,M$_{\odot}$) end their AGB phase with a larger fraction of carbon at their surface with respect to their lower-mass counterparts. This is due to the larger number of TDU episodes they experience before they lose their envelope entirely. In accordance with what was claimed in PV15, we conclude that, thanks to the good agreement found among different evolutionary codes in the final carbon surface mass fraction \citep{ventura18, fishlock14} and the small uncertainties in the C/H measurements, the carbon abundances provide strong constraints for the characterization of the progenitor mass. 

To retrieve the metallicity of each PN, we used the oxygen abundance and compared them with the theoretical abundances of different set of ATON models with different metallicities, as shown in  Fig. \ref{fco}.  We note that oxygen surface abundance is slightly enhanced during the TDU episodes experienced by LIMSs. Nevertheless, this variation is smaller (less than 0.1 dex) than the differences between the metallicities ($>$0.2 dex). The accuracy of the measured oxygen abundance is high enough to distinguish between the different metallicities, with the only exception being the SMP LMC 34 discussed in detail in Sect. \ref{SMP LMC 34sec}.   

Once the progenitor mass and the metallicity of each target were identified, we compared the luminosity and effective temperature derived from the SED modeling for each PN with the evolutionary tracks of the corresponding metallicity. This is shown in  Fig. \ref{tracks} where we plot the results of each PN in the Hertzsprung–Russell diagram. This comparison is fundamental for two reasons: (a) it highlights the consistency between the physical parameters derived from the SED analysis and the progenitor mass determined from the C and O measurements; and (b) it identifies which evolutionary track better reproduces the evolutionary history of each target, creating a link with previous AGB evolutionary history.

\begin{table}
\caption{ATON models of different metallicities (Z=8$\times 10^{-3}$, 4$\times 10^{-3}$, 2$\times 10^{-3}$, and 1$\times 10^{-3}$).  }
\label{tabatonab}      
\centering
%\addtolength{\leftskip}{+2cm}
\begin{tabular}{c c c c c c c c c}
\hline
Progenitor's mass [$\rm{M}_\odot$] & 12+log(C/H) & 12+log(O/H) \\
\hline
 & 8$\times 10^{-3}$ \\
\hline 
2.5  & 9.18  & 8.65 \\
2.0  & 8.87  &  8.62 \\ 
1.5  & 8.66  &  8.59 \\
%0.9  &  8.06 &   8.57 \\
\hline 
 & 4$\times 10^{-3}$ \\
\hline
1.5   & 8.79   & 8.33 \\
1.25 &  8.66  & 8.31  \\
1.0   &  8.62  & 8.31 \\
0.9   &  8.46  & 8.29 \\
0.8   &  7.73  & 8.26  \\
0.7   &  7.74  & 8.26  \\
0.6   &  7.72  &  8.26  \\
\hline
 & 2$\times 10^{-3}$ \\
\hline 
1.25   & 8.96   &   8.11  \\ 
1.0     & 8.59   &   8.05  \\
0.9     & 8.35   &   8.04  \\
0.85   & 8.20   &   8.02  \\ 
0.8     & 8.21   &   8.02  \\
\hline
 & 1$\times 10^{-3}$ \\ 
\hline 
%1.25   & 8.84  & 7.83 \\
0.8     & 8.70  & 7.84 \\ 
0.75   & 6.96  & 7.69 \\  
0.7     & 6.94  & 7.69 \\
\hline \\
\end{tabular} \\
\textbf{Notes.} The progenitor's mass is expressed at the beginning of 
the AGB phase; the abundances of carbon and oxygen 
are  those at the end of the AGB phase.
\end{table}

\begin{table}
\caption{Progenitor mass at the beginning of the AGB phase and the metallicity obtained for each PN in the sample studied.}
\label{tabmod}      
\centering
%\addtolength{\leftskip}{+2cm}
%\addtolength{\leftskip}{-0.5cm}
\begin{tabular}{c c c}    
\hline
ID & M$/\rm{M}_\odot$ & Z\\
\hline 
SMP LMC 4   & 1.5 & $8\times 10^{-3}$ \\ \\ 
SMP LMC 18  & 0.8$-$0.9 & $1-2\times 10^{-3}$ \\ \\ 
SMP LMC 25  & 0.9 & $2\times 10^{-3}$ \\ \\
SMP LMC 34  & 0.8 $-$ 0.9 & $2-4\times 10^{-3}$ \\ \\
SMP LMC 66  & 0.9$-$ 1.0 & $4\times 10^{-3}$ \\ \\
SMP LMC 71  & 2.0 & $8\times 10^{-3}$ \\ \\
SMP LMC 80  & $\lesssim 0.8$ & $4\times 10^{-3}$ \\ \\
SMP LMC 81  & $\lesssim 0.8$ & $4\times 10^{-3}$ \\ \\
SMP LMC 102 & 1.25 & $4\times 10^{-3}$ \\ \\
\hline
\end{tabular}
\end{table}

By collecting all the information derived by Figs. \ref{tracks} and \ref{fco}, we were able to determine the progenitor's mass of each object, with an uncertainty $\leq 0.5\,\rm{M}_\odot$. In Table \ref{tabmod} we report a summary of the progenitor's masses and the metallicities derived in the present work. In the following subsections we discuss the results obtained for each PN.

\subsection{SMP LMC 4 and SMP LMC 71: Carbon-rich, intermediate-mass stellar progenitors}
As illustrated in Fig. \ref{fco}, oxygen measurements indicate that SMP LMC 71 and SMP LMC 4 are characterized by a metallicity $\rm{Z}=8\times 10^{-3}$. Through our SED analysis, we have determined that the CSs have effective temperatures of 164\,000\,K and  105\,000\,K. These sources are the most luminous in the sample, with SMP LMC 71 reaching 5\,400$\,\rm{L}_\odot$ and SMP LMC 4 reaching 6\,500$\,\rm{L}_\odot$ (see Table \ref{tab1}). Thanks to their highest luminosity within the sample, their are placed in the upper region of the H-R diagram (left panel, Fig. \ref{tracks}), corresponding to progenitor's masses equal to $\sim 2.0\,\rm{M}_\odot$ for SMP LMC 71 and $\sim 1.5\,\rm{M}_\odot$ for SMP LMC 4. These estimates are in agreement with those derived using the carbon and oxygen abundances presented in Fig. \ref{fco}. This characterization suggests that SMP LMC 71 and SMP LMC 4 are intermediate-mass stars that experienced several TDU episodes, leading to enrichment in carbon on their surface and promoting the production of carbon dust. Additionally, SMP LMC 71 and SMP LMC 4 exhibit the highest value of the dust-to-gas mass ratio $\delta$[C] among the studied PNe (see Table \ref{tab1}), further supporting this interpretation.

\subsection{SMP LMC 25, SMP LMC 66, and SMP LMC 102: Carbon-rich, low-mass stellar progenitors}
SMP LMC 66 and SMP LMC 102 are two PNe with metallicity $\rm{Z}=4\times10^{-3}$ (see Fig. \ref{fco}). The characteristics of SMP LMC 102 have been extensively discussed in \cite{flavia23}, where a progenitor mass of $1.25\,\rm M_{\odot}$ was estimated. Consistent with this finding, our analysis of the SED reveals a luminosity of 3\,600\,$\rm{L}_\odot$ and an effective temperature  140\,000\,K (see Table \ref{tab1}). These parameters place SMP LMC 102 on the expected evolutionary track (see the left panel of Fig. \ref{tracks}). Additionally, the progenitor's mass of  1.25$\,\rm{M}_\odot$ for SMP LMC 102 is also consistent with the carbon and oxygen surface abundances derived from our models (see Fig. \ref{fco}). 

SMP LMC 66 lacks an observed HST/STIS spectrum, leading to larger uncertainty regarding its luminosity (L=4\,500$^{+2\,000}_{-1\,500}\,\rm{L}_\odot$). Based on the carbon and oxygen abundances shown in Fig. \ref{fco}, we suggest that SMP LMC 66 follows a similar evolutionary path to SMP LMC 102, with a progenitor mass range of $0.9-1.0\,\rm{M}_\odot$. Both SMP LMC 66 and SMP LMC 102 descend from a low-mass carbon star: a few TDU episodes allowed their progenitors to reach the carbon star (C/O>1) phase.

The metallicity of SMP LMC 25 is estimated to range between $2\times10^{-3}$ and $4\times10^{-3}$ based on the oxygen abundance in \citep{leisy06}. The most probable metallicity of SMP LMC 25 lies closer to the lower limit (i.e., $\rm{Z}\sim 2\times10^{-3}$). This inference is drawn from the discrepancy between a higher dust-to-gas ratio in SMP LMC 25 compared to SMP LMC 66 and SMP LMC 102, which would be inconsistent with a higher metallicity. If SMP LMC 25 had a metallicity closer to the upper limit, its progenitor mass would be around $\sim 0.8\,\rm{M}_\odot$ (see Fig. \ref{fco}), contradicting the observed high dust-to-gas ratio.
Considering a luminosity of 4\,900$\,\rm{L}_\odot$ and an effective temperature of 60\,000\,K derived from SED analysis, the PN progenitor mass for Z$=2\times 10^{-3}$ is estimated to be $\sim 0.85\,\rm{M}_\odot$, consistent with the carbon and oxygen abundances presented in Fig. \ref{fco}.

\subsection{SMP LMC 34: The lower limit to the carbon-rich stellar phase}
\label{SMP LMC 34sec}
SMP LMC 34 CS has a luminosity of 4\,500$\,\rm{L}_\odot$ and an effective temperature of 46\,000\,K. However, characterizing its metallicity poses a greater challenge compared to other PNe due to the significant uncertainty in the reported oxygen abundance (12+log(O/H)= $8.46^{+0.4}_{-0.4}$) as shown in Fig. \ref{fco} . 
Nevertheless, the presence of carbonaceous dust strongly suggests that SMP LMC 34 underwent a carbon-rich phase during its evolution. This, coupled with the measured carbon abundance, offers valuable constraints on the metallicity of this PN. 
In the AGB phase, stars with $\rm{Z}=8\times 10^{-3}$ that reach the carbon star stage typically have masses $\rm{M}>0.9\,\rm{M}_\odot$. These stars give rise to PNe with higher surface carbon abundances (12+log(C/H)>8.2) than what is measured for SMP LMC 34.  Conversely, stars with a masses $0.8\,\rm{M}_\odot < \rm{M} < 0.9\,\rm{M}_\odot$ within the metallicity range of $2\times 10^{-3} < \rm{Z} < 4\times 10^{-3}$ are known to reach the carbon star stage during the AGB phase. Their evolution exhibits luminosities, effective temperatures, surface carbon, and oxygen abundances compatible with the values measured for SMP LMC 34. Therefore, the most probable metallicity for SMP LMC 34 falls within the range  $2\times 10^{-3} < \rm{Z} < 4\times 10^{-3}$.

\subsection{SMP LMC 18: A late thermal pulse source}

According to its oxygen abundance, SMP LMC 18 have $10^{-3}< \rm{Z}<2\times 10^{-3}$.  From the SED analysis, we determined a CS luminosity of 2\,000$\,\rm{L}_\odot$ and an effective temperature of 50\,000\,K. SMP LMC 18 luminosity is the lowest within the sample, suggesting an oxygen-rich star as progenitor. However, the IR spectrum shows clear signatures of carbon dust presence (see the middle panels of Fig. \ref{f1}). In addition, the C/O>1 indicates that the progenitor mass has experienced a carbon-rich phase during its evolution. The combination of its low luminosity and the presence of carbon-rich nebular gas and dust renders SMP LMC 18 a peculiar object. A plausible explanation for this anomaly is that this PN derives from a star experiencing a late thermal pulse episode that have caused lower luminosity evolution. This scenario aligns with the evolutionary track  of a star of mass $0.9\,\rm{M}_\odot$ and a metallicity of $\rm{Z}=2\times 10^{-3}$, as shown in the right panel of Fig. \ref{tracks}. A progenitor with such characteristics would be consistent with the observed abundances of SMP LMC 18 (see Fig. \ref{fco}).

\subsection{SMP LMC 80 and SMP LMC 81: The oxygen-rich sources}
\label{oxygensec}
SMP LMC 81 and SMP LMC 80 exhibit C/O<1. The metallicity, derived from the comparison of the observed abundances and the surface abundance at the end of the AGB phase computed using the ATON code, is $\rm{Z}=4\times10^{-3}$ (see Fig. \ref{fco}). From the SED analysis, we find their luminosities to be 4\,700$\,\rm{L}_\odot$ and  3\,200$\,\rm{L}_\odot$, and that the effective temperatures are 80\,000\,K and 57\,000\,K, respectively. These values position them in the lower part of the H--R diagram (right panel of Fig. \ref{tracks}), indicative of progenitor masses $\lesssim0.8\,\rm{M}_\odot$.  In this mass range, the envelope thickness is so low that a small variation of carbon in its surface could result in a significant difference in the final C/O ratio. Consequently, the derived progenitor's mass is subject to uncertainty.

\section{Dusty features in the sample}
\label{dust}
The IR spectra of the seven CRD PNe in our sample show the presence of amorphous carbon, graphite and, some sources, unidentified IR features, fullerenes, as well as emissions attributable to SiC and PAHs, among others (see Appendix \ref{appendix}). In \cite{sloan14}, PAHs are grouped into three different classes. The class B PAHs, which \cite{peeters02} claimed to be the most frequent in the PNe, are characterized by an emission at $7.85~\mu$m \citep{sloan14}. In accordance with that, we find that most of our CRD PNe present signature that can be connected to the presence of class B PAHs. This is the case of SMP LMC 25 and SMP LMC 71, already studied for their PAH features in \citet{stanghellini07}. Despite their noisy \textit{Spitzer} spectra at wavelength $5.5<\lambda[\mu \rm{m}]<13$, we suggest that SMP LMC 18 and SMP LMC 34 could also present some traces of PAHs emission. We find that by adding the contribution from PAHs to the modeling, we are able to achieve a better description of the photometric point at 8$\,\mu$m (see Appendix \ref{appendix} for more details). Furthermore, as already claimed by \citet{sloan14}, in SMP LMC 25 we observe the presence of the $6.2\,\mu$m PAH feature, which could explain the detachment of the photometric filter at $5.6\,\mu$m (see Fig. \ref{f1}). We believe that SMP LMC 71 could also present the same emission, which would help fill the gap between the synthetic and the observed photometry at $5.6\,\mu$m (see Fig. \ref{f2}). Three sources (SMP LMC 18, SMP LMC 25, and SMP LMC 34) exhibit an unidentified broad $26-30\,\mu$m feature ({the 30\,$\mu$m feature), which could be explained by MgS dust \citep[for details, see][]{sloan14}; for the same objects we also observe another unidentified broad IR feature around $15-20~\mu$m (for the details of each PN and the comparison with the literature, see Appendix \ref{appendix}). 

In the SED of five CRD PNe (SMP LMC 4, SMP LMC 18, SMP LMC 25, SMP LMC 66, and SMP LMC 71), we highlight the presence of a near-IR bump spanning wavelengths $3\lesssim \lambda [\mu m] \lesssim 8$. This peculiar feature cannot be reproduced if the dust temperature is determined self-consistently within the photoionization model. To characterize the near-IR bump, we conducted a two-step process: in the first iteration, we find the parameters of the CS, nebula, and dust that best reproduce the SED (as reported in Sect. \ref{input}), allowing the photoionization model to find the dust temperature in a self-consistent manner.  The resulting synthetic spectrum reproduces the observed SED in the UV, in the optical and in the mid-IR, with the exception of the near-IR bump. The dust temperatures of this model range between $85$ and $135$\,K. Next, we ran a second model assuming all the parameters used in the previous run, but imposing a temperature that better agrees with the near-IR photometry and assuming a pure amorphous carbon mineralogy. This second model indicates that the near-IR bump is compatible with a "hot dust" component with temperatures ranging between $600$ and $800$\,K.

SMP LMC 81 shows traces of silicate dust and, in agreement with LS07, we highlight the presence of a features at $\sim$9.8 and 18\,$\mu$m probably due to amorphous silicates. Despite the absence of clear silicate features, the presence of a small amount of silicate dust in the PN SMP LMC 34 nebula cannot be excluded. 

Due to the almost negligible IR emission detected in the \textit{Spitzer} spectrum, SMP LMC 80 shows no evidence of dust emission. Given the low-mass progenitor of the CS of SMP LMC 80 (see Sect. \ref{oxygensec}), a low dust production is expected \citep{ventura14,flavia15}. The current $\rm{T}_{\rm{eff}}$ is much higher than those typical of the AGB phase when the dusty layer forms; therefore, we suggest that the already low amount of dust formed during its evolution could have been completely destroyed in the PN phase.

\section{Conclusions}
In this study we present a multiwavelength investigation of nine round and elliptical PNe in the LMC, with the aim of understanding the dust production of LIMSs. Utilizing spectra, photometric data, and abundances from the literature, we conducted a SED analysis from the UV to the IR through the use of photoionization models. We characterized various physical parameters, including the CS luminosity and effective temperature, the radial dimension and gas mass of the nebulae, the chemical composition of the dust, and the dust-to-gas ratio. Our primary goal was to connect these findings to the sources' past evolutionary history. To achieve this, we computed a wide grid of ATON evolutionary tracks extended to the PN phase. We derived the metallicity of each PN by comparing observed carbon and oxygen abundances with the ATON predictions. Furthermore, we compared ATON evolutionary tracks with the luminosity and effective temperature derived from SED analysis. This combination of abundance information and evolutionary tracks enabled us to determine the PN progenitor masses and metallicity.

Of the studied PNe, two with intermediate-mass progenitors (1.5--2.0\,$\rm{M}_{\odot}$) exhibit a significantly  higher dust-to-gas ratio compared to the rest of the sample, which evolved from lower-mass progenitors (0.8-1.25\,$\rm{M}_{\odot}$). Specifically, five carbon-rich PNe (SMP LMC 4, SMP LMC 18, SMP LMC 25, SMP LMC 66, and SMP LMC 71) exhibit a near-IR bump peaking around $3\lesssim \lambda [\mu m] \lesssim 8$. This near-IR feature, distinct from the longer wavelength emission attributed to cool dust, suggests the presence of "hot" dust with temperatures around 600--800\,K. 

In a future paper, we intend to use the present findings to explore the mass-loss history and dust-production rate during the AGB phase. Our investigation will focus on establishing the relationship between the AGB-to-PN transition timescale and the dust-to-gas ratio as a function of the CS masses and metallicities.

\begin{acknowledgements}
F.D.A is supported by the INAF-Mini-GRANTS 2023 "Understanding evolved stars and their dust production through the lens of planetary nebulae". DK acknowledges the support of the Australian Research Council (ARC) Discovery Early Career Research Award (DECRA) grant (DE190100813) and this research was supported in part by the Australian Research Council Centre of Excellence for All Sky Astrophysics in 3 Dimensions (ASTRO 3D) through project number CE170100013. MAGM and DAGH acknowledge the support from the State Research Agency (AEI) of the Spanish Ministry of Science and Innovation (MCIN) under grant PID2020-115758GB-I00/AEI/10.13039/501100011033. This article is based upon work from European Cooperation in Science and Technology (COST) Action NanoSpace, CA21126, supported by COST. This work is based on observations made with the \textit{Spitzer} Space Telescope, which was operated by the Jet Propulsion Laboratory, California Institute of Technology under a contract with NASA.

\end{acknowledgements}


\begin{appendix}

\section{Details of the photoionization models for the sample PNe}
In the following paragraphs we present the most relevant information available in the literature regarding the photometry, UV, and IR spectra that we used for the present investigation. When available, we compare the effective temperatures with the ones obtained by EV03 using the Zanstra method. Furthermore, we describe the scheme of the synthetic models that we used to perform the SED analysis, whose input are reported in Table \ref{tabinput} and the main output in Table \ref{tab1}. The best set of parameters was obtained by running CLOUDY repeatedly until the best agreement with the observational data was reached (i.e., the photometry and the UV and IR spectra). We note that for a few PNe, the HST and \textit{Spitzer} spectra are noisy; this makes the modeling more challenging as we have to rely mainly on the photometric data.

\label{appendix}

\subsection*{A.1. SMP LMC 4}
SMP LMC 4 is a moderately high excitation nebula (LS05) characterized by an elliptical shape and a faint halo (LS07). The HST spectrum is quite noisy in the region between 1\,500\,$\AA$ and 1\,732\,$\AA$ and shows an emission line of C III] at 1\,908\,$\AA$. The \textit{Spitzer} spectrum presents nebular atomic emission lines that are characteristic of a very high excitation (LS07). We distinguish the [OIV] line at 26\,$\mu$m and very weak emission at 11.3\,$\mu$m, which could be of interstellar nature or the remnant of evaporated dust (LS07; see the upper panel of our Fig. \ref{f1}).

We find an $\rm{T}_{\rm{eff}}=$105\,000$^{+2\,000 }_{-5\,000 }$\,K, in good agreement with the Zanstra temperature (HeII) reported in EV03 ($\rm{T}_{eff}^{HeII}=$89\,900$\pm 7\,200$\,K). For the dust, LS07 claimed a featureless spectrum, which we modeled using only amorphous carbon, obtaining a dust temperature $\rm{T}_d=85^{+20}_{-19}$\,K. SMP LMC 4 is the only case in our sample where the dust temperature self consistently calculated by CLOUDY is too low to reproduce the IR peak (see Fig. \ref{f1}). By imposing a dust temperature, the best agreement with the observational data is obtained  if $\rm{T}_d=150$\,K and log($\delta$[C])= $-3.27 ^{+0.04} _{-0.09}$. Furthermore, to get the mid-IR photometry, it is necessary to add a near-IR bump composed of amorphous carbon, characterized by a temperature $\rm{T}_d=630$\,K (the details on the near-IR bump are reported in Sect. \ref{dust}). The electron temperature is $1.24\times 10^4<\rm{T}_e\rm{[K]}<1.35\times 10^4$.

\subsection*{A.2. SMP LMC 18}
SMP LMC 18 is a low-excitation nebula (LS05) characterized by a round shape (LS07). The HST spectrum is quite noisy in the region between 1\,500\,$\AA$ and 1\,732\,$\AA$ and shows an emission line of C III] at 1\,908\,$\AA$. Furthermore, LS07 claimed the possible presence of the broad $15-20$ and $30\,\mu$m features (see the middle panel of Fig.\ref{f1}).

The effective temperature of SMP LMC 18 is found to be $\rm{T}_{\rm{eff}}=$50\,000  $^{+6\,000 }_{-5\,000 }$\,K. Since this value of effective temperature is outside of the Rauch atmosphere grid \citep{rauch03}, we used the \cite{pauldrach01} models of atmosphere. Using the \cite{pauldrach01} parameters, we obtain a value of effective temperature close to the one reported in LS05, which was calculated from the UV continuum using a black body ($\rm{T}_{\rm{eff}}^{\rm{BB}}\sim$ 40\,000\,K). Concerning the dust continuum, LS07 claimed a weak featureless spectrum, which we modeled using amorphous carbon and graphite. Despite the absence of any signature of the PAHs in the noisy \textit{Spitzer} spectrum, we added their contribution to better reproduce the photometric point an 8$\,\mu$m, finding that the best description is achieved using log($\delta$[PAH])=$-5.04^{+0.22} _{-0.38}$. Furthermore, to reproduce the mid-IR photometry, it is necessary to add a near-IR bump composed of amorphous carbon, characterized by a temperature $\rm{T}_d=500$\,K (see Sect. \ref{dust} for the detail on the near-IR bump modeling). Finally, we find $1.17\times 10^4<\rm{T}_e\rm{[K]}<1.21\times 10^4$. 

\subsection*{A.3. SMP LMC 25}
SMP LMC 25 is an intermediate-excitation nebula (LS05) characterized by a round shape (LS07). The HST spectrum shows strong emission lines of C III] at 1\,908\,$\AA$ and C II] at 2325--2329\,$\AA$ (LS05). The \textit{Spitzer} spectrum shows peaks at 6.2 and 7.7$\,\mu$m, all likely to be proto-PAH features, and a SiC emission at 11.3$\,\mu$m (LS07).  From the IR is possible to recognize some of the fullerene emission (mainly due to C$_{60}$) at 7.0, 8.5, 17.4 and 18.9\,$\mu$m \citep{anibal} and a broad unidentified 15–20\,$\mu$m feature that could be due to relatively large PAHs or PAH clusters. SMP LMC 25 was also studied by \cite{sloan14}, who confirmed the presence of fullerene and a structure between 6 and 9$\,\mu$m usually associated with PAHs. We highlight the presence of two further bumps, the first centered at $15-20~\mu$m and the second at 30$\,\mu$m; the latter is probably due to the MgS dust \citep{sloan14}.

We adopted a dust schematization based on amorphous carbon, graphite, and SiC, which is consistent with the carbon dust feature observed in SMP LMC 25 claimed by LS07.  We also added the contribution of the PAH, with log($\delta$[PAH])=$-3.77^{+0.30} _{-0.10}$, in agreement with \citet{anibal2}. Due the presence of a strong emission at $6.2\,\mu$m \citep[see the lower panel of our Fig. \ref{f1}]{sloan14}, we believe that different optical constants of PAHs (not presently available in CLOUDY) could explain the difference between our synthetic photometric data at $5.6\,\mu$m and the observed data. Furthermore, to reproduce the mid-IR photometry it is necessary to add a near-IR bump composed of amorphous carbon, characterized by a temperature $\rm{T}_d=780$\,K (the details on the near-IR bump are reported in Sect. \ref{dust}). Concerning the electron temperature we find $7.34\times 10^3<\rm{T}_e\rm{[K]}<1.41\times 10^4$ .

\subsection*{A.4. SMP LMC 34}
SMP LMC 34 is a low-excitation nebula (LS05) characterized by an elliptical shape (LS07). The HST spectrum is quite noisy in the region between 1\,500\,$\AA$ and 1\,732\,$\AA$ and shows an emission line of C III] at 1\,908\,$\AA$. The IR spectrum shows the broad unidentified 30$\,\mu$m feature that could be linked to the presence of MgS dust \citep{sloan14} and the emission of [NeIII] at 15.6\,$\AA$ (see the upper panel of Fig. \ref{f2}). 

The effective temperature is found to be $\rm{T}_{\rm{eff}}=46\,000  ^{+4\,000 }_{-3\,000 }$\,K. Being lower than 50\,000\,K, \cite{pauldrach01} atmospheres are adopted for the modeling. The effective temperature derived in the present work is lower than the measurement reported in EV03, calculated with the Zanstra method for the HeII ($\rm{T}_{eff}^{HeII}=$67\,800$\pm 3\,600$\,K). That divergence on the effective temperature calculated from the UV continuum fitting was already highlighted in LS05, which suggested a lower value for the effective temperature, around 40\,000\,K, in better agreement with our results. To reproduce the IR continuum, we used a combination of amorphous carbon and graphite, consistent with the featureless spectrum claimed by LS07. We note that the \textit{Spitzer} spectrum is too noisy in the wavelength $5.5<\lambda[\mu \rm{m}]<13$ to confirm or discard the presence of PAHs. Their contribution is added to better reproduce the photometric data at 8$\,\mu$m, finding log($\delta$[PAH])=$-4.89^{+0.28} _{-0.23}$. Given the large error bar on the oxygen abundance (12+log(O/H)= $8.46^{+0.4}_{-0.4}$), a C/O$\sim$1 or slightly lower than unity cannot be excluded. Therefore, we tried to perform the SED analysis by adding the presence of silicate grains, residual of the dust formed during the oxygen-rich phase, to the previous dust mixture (amorphous carbon, graphite, and PAHs). From this exploration, we find that the goodness of the models is still reliable by assuming $\delta$[C]=$2.24\times 10^{-4}$ and $\delta$[Sil]=$4.04\times 10^{-5}$ and therefore we cannot rule out the possibility that the dust of SMP LMC 34 is characterized by a mixed chemistry. 
On the contrary, by assuming a pure amorphous silicate dust we obtain a synthetic spectrum in disagreement with the \textit{Spitzer} spectrum and most of the observed IR photometry. For the electron temperature, we derive $8.44\times 10^3<\rm{T}_e\rm{[K]}<1.30\times 10^4$.

\subsection*{A.5. SMP LMC 66}
SMP LMC 66 is a high-excitation nebula characterized by an elliptical shape (LS07). The \textit{Spitzer} spectrum is dominated by the emission lines of [SIV] at 10.3 $\mu$m, [NeIII] at 15.6 $\mu$m and [OIV] at 26 $\mu$m (see Fig. \ref{f3}). We highlight the absence of the HST spectrum and of the V magnitude because of the bad quality flag of the measurement of \cite{zaritsky04}.

To reproduce the emission due to the dust, LS07 claimed a featureless spectrum, which we modeled using amorphous carbon and graphite. We find a discrepancy between the observed photometric data at $8~\mu$m with respect to the synthetic one. We believe that the PAH emission at 7.85$\,\mu$m, which can be recognized in most of the PNe \citep{sloan14}, could help fill the gap in that spectral region (see the upper panel of Fig. \ref{f3}). This nebula shows a near-IR bump reproduced by including amorphous carbon with temperature $\rm{T}_d=550$\,K (see Sect. \ref{dust} for the detail on the near-IR bump modeling). Concerning the electron temperature we find $1.28\times 10^4<\rm{T}_e\rm{[K]}<1.46\times 10^4$.

\subsection*{A.6. SMP LMC 71}
SMP LMC 71 is a high-excitation nebula characterized by an elliptical shape (LS07). The HST spectrum shows emission lines of C III] at 1\,908\,$\AA$ (see the middle panel of Fig. \ref{f2}). By looking at the UV spectrum, LS05 claimed a moderately high-excitation nature of the nebula but another scenario appears evident by looking at the \textit{Spitzer} spectrum, in which the nebula results with a very high excitation (LS07). In the IR spectrum, there is a forest of emission lines in which it is possible to recognize a broad plateau unidentified feature at 6--9$\,\mu$m, possibly due to small grain clusters, superimposed on narrow features at 6.2, 7.7, and 8.6$\,\mu$m, characteristic of classical PAHs (LS07). In agreement with LS07, we see a possible 30 µm feature (see the middle-right panel of Fig. \ref{f2}) that could be due to MgS dust \citep{sloan14}. We highlight the absence of the V magnitude because of the bad quality flag of the measurement of \cite{zaritsky04} (see the middle-left panel of Fig. \ref{f2}). 

The effective temperature is found to be $\rm{T}_{\rm{eff}}=$164\,000 $^{+9\,000 }_{-4\,000 }$\,K, higher than the one obtained from the Zanstra method for the He II given in EV03 ($\rm{T}_{eff}^{HeII}=$83\,400$\pm 5\,200$\,K), but still reasonable since their measurement is based on a lower limit in the V magnitude (see Table 3 of EV03). For the dust, LS07 claimed the presence of carbon dust features superimposed on the dust continuum emissions, in agreement with our description based on amorphous carbon, graphite, PAHs, and so on.  We believe that our choice of optical constants is not the best to reproduce the central peak of the dust. Indeed, to achieve the photometric point at 5.6$\,\mu$m it would be necessary to use new optical constants with respect to the CLOUDY built-in that we adopted. In particular, we believe that the PAH emission at $6.2\,\mu$m, which can be recognized, for example, in SMP LMC 25 \citep{sloan14}, could help fill the gap in that region (see the upper panel of Fig. \ref{f3}). However, using the mineralogy described above we find log($\delta$[C])=$-2.21 ^{+0.15} _{-0.10}$ and log($\delta$[PAH])=$-3.46 ^{+0.09} _{-0.19}$. This nebula shows a near-IR bump reproduced by including amorphous carbon with temperature $\rm{T}_d=680$\,K (the details on the near-IR bump are reported in Sect. \ref{dust}). As a result of the SED analysis, we also find $1.4\times 10^4<\rm{T}_e\rm{[K]}<1.5\times 10^4$.

\subsection*{A.7. SMP LMC 80}
SMP LMC 80 is a low excitation nebula (LS05) characterized by a round shape (LS07). The HST spectrum is quite noisy in the region between 1\,077\,$\AA$ and 1\,732\,$\AA$. The \textit{Spitzer} spectrum shows intermediate-excitation emission lines (LS07) with a strong emission of [NeIII] at 15.6\,$\mu$m and [SIII] at 18.7\,$\mu$m (see the middle panel of Fig. \ref{f4}). 

We cannot reach an agreement with most of the photometric data, maybe because of a ring shape, visible in [NII] \citep{shaw01}. For the dust, we confirm the featureless spectrum claimed by LS07, modeling the \textit{Spitzer}/IRS spectrum with a negligible amount of amorphous silicate dust. Furthermore, we find $1.25\times 10^4<\rm{T}_e\rm{[K]}<1.34\times 10^4$.

\subsection*{A.8. SMP LMC 81}
SMP LMC 81 is an intermediate optical excitation nebula (LS05) characterized by a round shape (LS07). The HST spectrum is quite noisy in the region between 1\,077\,$\AA$ and 1\,732\,$\AA$ and shows emission lines of C III] at 1\,908\,$\AA$ and HeII at 1\,640\,$\AA$. The \textit{Spitzer} spectrum shows nebular emission of intermediate excitation lines (LS07) and distinct amorphous silicate features at $\sim$9.8 and 18$\,\mu$m (see the upper panel of Fig. \ref{f4}). Using the built-in CLOUDY optical constants, we cannot perfectly model the amorphous silicate features; especially the one at $\sim9.8\,\mu$m. This aspect, however, does not affect the determination of the main parameters of the CS, the nebula and the dust, which is the main goal of the present work.

Concerning the dust, LS07 claimed an ORD chemistry, which we modeled using amorphous silicates. From the SED analysis, we also derive $8.28\times 10^3<\rm{T}_e\rm{[K]}<1.61\times 10^4$.

\subsection*{A.9. SMP LMC 102}
SMP LMC 102 is a high-excitation nebula (LS05) characterized by a round shape (LS07). The HST spectrum is dominated by the emission lines of CIV at 1\,548\,$\AA$, HeII at 1\,640\,$\AA,$ and C III] at \,908\,$\AA,$ whereas the \textit{Spitzer} spectrum has a prominent emission line due to [OIV] at 26\,$\mu$m (see the lower-right panel of Fig. \ref{f1}). On the photometric point of view, the measurement of \textit{Spitzer} and the U, B, V data from \cite{reid14} (see the lower panel of Fig. \ref{f1}) were not available. 

The effective temperature is $\rm{T}_{\rm{eff}}=$140\,000$^{+2\,300}_{ -2\,000 }$\,K, which is in agreement with the ones derived from the Zanstra method for the He II from EV03 ($\rm{T}_{eff}^{HeII}=$131\,800$\pm 12\,400$\,K). For the dust, LS07 claimed a featureless spectrum, which we modeled using only amorphous carbon. As an output parameter, we also derive an electron temperature $1.39\times 10^4<\rm{T}_e\rm{[K]}<1.71\times 10^4$.

\end{appendix}

\end{document}